\documentstyle[12pt]{article}
\date{25.11.97}
\begin{document}

\title{Covariant Quark Model for the Baryons}

\author{F. Coester$^1$, K.Dannbom$^2$ and D.O. Riska$^2$}
\maketitle

\centerline{\it $^1$Physics Division, Argonne National Laboratory,
Argonne IL 60439-4843, USA}

\centerline{\it $^2$Department of Physics, University of Helsinki,
00014 Finland}

\setcounter{page} {0}
\vspace{1cm}

\centerline{\bf Abstract}
\vspace{0.5cm}

A family of simply solvable covariant quark models for the baryons is
presented. With optimal parameter choices the models reproduce the
empirical spectra of the baryons in all flavor sectors to an accuracy
of a few percent. Complete spectra are obtained for all states of the 
strange, charm and beauty hyperons with 
$L \leq 2$. The magnetic moments and axial
coupling constants of the ground state baryons correspond to
those of conventional quark models. We construct current-density 
operators that are consistent with empirical nucleon form factors at low 
and medium  energies.

\newpage
\centerline{\bf 1. Introduction}
\vspace{0.5cm}

A theoretically consistent description of the baryons based on the
concept of constituent quark should build in confinement 
and Poincar\'{e} invariance from the beginning. Confinement is an
obvious consequence of the unbounded discrete mass spectrum. 
The requirements of Poincar\'e  invariance for the mass
operator are easily met by explicit construction: The Hilbert space
of states ${\cal H}$ is the tensor product,  
${\cal H}={\cal H}_C\times {\cal H_\ell}$, of functions 
representing the center-of-mass motions with the ``little Hilbert space''
\cite{RHa} which is the representation space of the ``little group''.
The eigenfunctions of the four-momentum obtained with such
a mass operator depend on a choice of boosts and kinematic variables.
These eigenfunctions are related to observables
only in combination with a
corresponding representation of the 
current density operators. The
construction of a Poincar\'e covariant quark model of the  baryons 
thus
involves both the construction of a mass operator with appropriate spectral
properties and the construction of Poincar\'e covariant current density
operators. For confined quarks the conventional assumption of a
free-quark
density is not compelling at low energies.
 \cite{CoRi}. Giving up that assumption 
greatly facilitates the construction of simple phenomenological
current operators that satisfy all symmetry requirements.
In this context instant-form kinematics appears to have two principal
virtues: (i) It allows recovery of the familiar features of conventional
quark models. In particular it facilitates 
implementation of the physical notion
of ``impulse'' currents. (ii)
It arises naturally when quantum mechanical models are derived from
Euclidean Green functions.\cite{Glim}

Here we consider confining mass operators which 
 provide satisfactory descriptions of
the baryon spectrum in all flavor generations. The mass operators
considered are sums of a confining term, 
which depends only on the Jacobi coordinates
of the 3-quark system, and a flavor and spin dependent
hyperfine correction. The former provides the basic shell structure of
the spectrum and is readily diagonalized by means of hyper-spherical
harmonics \cite{FaRi}. 
The latter is built on the observation that a superior
description of the baryon spectrum in all flavor generations may be
achieved by assuming that the main hyperfine correction should have
the flavor-spin structure 
--$\sum_{i<j}\vec{\lambda}_i\cdot \vec{\lambda}_j
\vec{\sigma}_i\cdot 
\vec \sigma_{j}$, which e.g. is suggested by the short
range part of the Goldstone boson exchange interaction between the
constituent quarks \cite{GloRis, GloRis1}. 

The aim of this work is to provide a class of simple models
that can account for all states in the
$S,P$ and $SD$ -shells of the baryon spectra in all flavor
generations. The parameter values in the models are extracted 
from the lowest splittings in the spectra of the
successive flavor generations. The motivation
for this work is the rapid progress in experimental 
determination of the low lying levels in the $C = +1$ charm  hyperon   spectra,
and the parallel - if somewhat slower - progress in the case of
the corresponding 
$B=-1$ hyperon states. We also give energy levels for the $C = 2,3$
hyperons, which are mainly  determined by the spin- and
flavor independent mass operator with only small
hyperfine corrections. These spectra should
give fairly direct information on the effective quark confining 
interaction. 

Quark currents  consistent with the covariance conditions, 
lead to values for the magnetic
moments and axial coupling constants of the ground state
baryons  similar to those obtained by
conventional quark models. The 3-quark wave functions that yield 
satisfactory baryon spectra have  small radii. These models
therefore require significant quark-charge
distributions to account for the observed nucleon form factors. 
With appropriate quark-charge  and current distributions 
the model current operators are consistent with
observed nucleon form factors at low and medium energies.
Applications to transitions are beyond the scope of the present paper.

This paper is divided into 8 sections. In section 2 we describe the
model  mass operator and the parameter choices. In section 3 we
derive nucleon form factors as well as  magnetic moments and axial coupling
constants of the ground-state baryons. The  current operators used are
 consistent with the Poincar\'e
representation specified by the mass operator and with instant-form 
kinematics.
The spectra of the nucleon and 
$\Delta$-resonances are described in section 4,
the spectra of the 
strange hyperons in section 5. Sections 6 and 7 contain 
predictions for the charm and beauty hyperons respectively. 
Section 8 contains a summary. 

\vspace{0.5cm}
\centerline{\bf 2. The Mass operator}
\vspace{0.5cm}

\centerline{2.1 The Hilbert Space of 3-Quark States }
\vspace{0.5cm}

The states in the baryon spectrum
are described by  vectors in the little Hilbert space 
${\cal H}_\ell$, which is 
the representation 
space of the direct product of
the little group, $SU(2)$, with flavor  and color $SU(3)$.
Concretely these states are
realized by functions, 
$\phi$,
of three quark positions $\vec r_i$, three spin variables $\mu_i$
and three flavor variables $f_i$, 
which are symmetric under permutations
and invariant under translations $\vec r_i\to \vec r_i+\vec a$. 
The translational invariance is realized by
expressing the wave functions  in terms of 
Jacobi coordinates
$$ \vec r:={1\over \sqrt{2}}(\vec r_1-\vec r_2),\quad
\vec \rho:=\sqrt{{2\over 3}}(\vec r_3-{\vec r_1+\vec r_2\over 2}).
\eqno (2.1)$$
Representations of the full Poincar\'e group
obtain on the 
tensor product, ${\cal H}:={\cal H}_\ell\otimes {\cal H}_c$, of 
the little Hilbert space
${\cal H}_\ell$ with the Hilbert space ${\cal H}_c$ 
of functions of the four-velocity $v$, which is 
specified by 3 independent components. Translations are generated
by the four-momentum operator $P={\cal M} v$. 
Any confining self-adjoint
mass operator ${\cal M}$, independent of $v$,  satisfies  all 
relativistic symmetry requirements if it is invariant under rotations. 
At the level of spectroscopy
alone there is no difference
between relativistic and 
nonrelativistic quark models because the Galilean rest energy
operator
satisfies all the symmetry requirements of  a mass operator.

The Poincar\'e invariant inner
product of the functions  representing baryon states is defined as
$$
(\Psi,\Psi)=\int d^4v 2
\delta(v^2+1)\,\theta(v^0)\int d^3\kappa\int
d^3k|\Psi(v,\vec \kappa,\vec k)|^2,
\eqno (2.2)$$
where $\vec \kappa$ 
and $\vec k$ are  the momenta conjugate to $\vec\rho$ and $\vec r$ .
Summation over spin and flavor variables is implied.
Under Lorentz transformations $v\rightarrow \Lambda v$ the vectors
$\vec\kappa$ and $\vec k$ and the three quark spin variables $\mu_1,\mu_2$, 
$\mu_3$ undergo Wigner rotations $R_W(\Lambda,v)$:
$$ R_W(\Lambda,v):= B^{-1}(\Lambda v)\Lambda B(v)\; .
\eqno (2.3)$$
Note that with 
canonical boosts, $R_W(\Lambda_v,v)=1, $
for any rotationless
Lorentz transformation $\Lambda_v$ in the direction of $\vec v$. 

For the construction of quark 
currents it is convenient
to define internal four-momenta $p$ and $q$ by
$$ p:=B(v)\{0,\vec \kappa\},\quad 
q:=B(v) \{0,\vec k\},\eqno(2.4) $$
so that
$p^2=|\vec \kappa|^2$  and $q^2=|\vec k|^2$. 

The construction of unitary representations of the Poincar\'e group   
sketched here naturally leads
to point-form kinematics. Once the eigenfunctions of the mass operator 
are known, it is easy
to realize unitary transformations to other forms of kinematics
explicitly \cite {CoRi}.
Let $\Psi_n(v,\vec \kappa,\vec k)$ be 
eigenfunctions of ${\cal M}$, with eigenvalues 
 $M_n$.
Any state  $\Psi=\sum_n \Psi_n c_n$ can be represented by functions 
$\Psi(\vec P,\vec p, \vec q)$
normalized as

$$(\Psi,\Psi)=\int d^3 P\int d^3 p\int d^3q |
\Psi(\vec P, \vec p,\vec q)|^2\; .
\eqno(2.5).$$
This representation is used below to calculate magnetic
moments and axial coupling constants.
The unitary transformation 
$\Psi(v,\vec \kappa, \vec k)\to \Psi(\vec P,\vec p, \vec q)$ 
is specified by the 
variable transformation  $\{v,\vec \kappa, \vec k, n\} 
\to \{\vec P,\vec p, \vec q,n\}$
where $\vec p =\vec p(v,\vec \kappa)$ and $\vec q 
=\vec q(v,\vec k)$ are specified by
eq. (2.5)
and $ \vec P = M_n\vec v$ in each term of the sum over $n$. 
 Individual quark
momenta are defined by
$$\vec p_1={1\over 3} \vec P-{1\over 2}
\sqrt{{2\over 3}}\vec p
+\sqrt{{1\over 2}}\vec q,$$
$$\vec p_2={1\over 3} \vec P-{1\over 2}
\sqrt{{2\over 3}}\vec p-\sqrt{{1\over 2}}\vec q,$$
$$\vec p_3={1\over 3} \vec P+\sqrt{{2\over 3}}\vec p\; ,
\label{SQP} \eqno(2.6)$$
so that $\vec P=\vec p_1+\vec p_2+\vec p_3$.

\vspace{1cm}

\centerline{2.2. The Confining Mass Operator}
\vspace{0.5cm}

We shall consider the confining  mass operators of the
form

$$ {\cal M}_0 =\sqrt{3 \{
\vec k^2 + \vec \kappa^2+f(R)\}+{\cal K}}, \eqno(2.7)$$
where   $R$,
$$
R:=\sqrt{2(\vec r^2+\vec \rho^2)}\; .
\eqno(2.8)$$
is the radius of a 6-dimensional hyper-sphere,
and the operator  ${\cal K}$  
$${\cal K}:=n_S\Delta_S^2+n_C\Delta_C^2+n_B\Delta_B^2,
\eqno (2.9)$$
represents the flavor dependence.
The integers $n_S$, $n_C $ and $n_B$ are flavor quantum numbers of
strangeness, charm and beauty. The parameters $\Delta_f$
 represent  the flavor dependence of the ground-state  baryon masses.
Mass operators of the form (2.7) 
satisfy all symmetry requirements and are
easily diagonalized by means of hyper-spherical harmonics.\cite{FaRi}

It is convenient to express the magnitudes $r$ and $\rho$
as functions of the radius $R$ and the auxiliary variable z,
$$
z:= {2(r^2-\rho^2)\over  R^2}\; ,
\eqno(2.10)
$$
so that
$$
r={R \over 2} \sqrt{1+z},\hspace{0.5cm} \rho ={ R \over 2}
\sqrt{1-z}\;.
\eqno(2.11)
$$
The eigenfunctions, $\phi(\vec{r},\vec{\rho})$, of the mass operator 
${\cal M}_0$ are  products,
$$
 \phi(\vec{r},\vec{\rho})=  u_{nK}(R)\;{\cal Y}_{K}   
\eqno(2.12)$$ 
of radial functions $u_{nK} (R)$ and the 
hyper-spherical  harmonics ${\cal Y}_K$,

$${\cal Y}_K=[Y_{\ell_1}(\hat {r}) \otimes Y _{\ell_2} (\hat \rho)]
_{\ell m}(1+z)^{\ell_1/2} (1-z)^{\ell_2/2} P_\nu 
^{\ell_1 + {1 \over 2},\ell_2 + {1 \over 2}} (z)\; , \eqno (2.13)$$
where $K = 2\nu + \ell_1+\ell_2 $ and $ P_\nu $ is a Jacobi
polynomial. The functions $ u_{nK} (R) $ are solutions to the 
radial equation
$$ \biggl[ 2 \biggl\{ - {1\over R^{5/2}} {d^2\over dR^{2}} R^{5/2}
 + {{\cal L}({\cal L} +1)\over
R^2} \biggr\}  + f (R) \biggr]  u_{nK} (R)= 
\epsilon^{2}_{nK} u_{nK} (R), \eqno(2.14)$$
where $ {\cal L}= K+ {3\over 2}$ is the orbital angular momentum and
 $n$ is the number of nodes in the radial wave function.
The normalization
condition for the
radial functions $ u_{nK} (R) $ are    
$${\pi \over 128} \int ^{\infty}_{o} dR R^{5} u^{2}_{nK}(R)=1.
\eqno(2.15)$$
The eigenvalues of the mass operator ${\cal M}_0$ are then

$$ {\cal E}_0 = \sqrt{\epsilon_{nK}^2+{\cal K}} .\eqno (2.16)$$

In the following we  consider  two models for the function $ f(R) $
in (2.7):  

$$  f_1(R) ={\omega^{4}\over 2} R^{2},\eqno(2.17)$$
and
$$ f_2 (R) = - {a\over R} + b R.\eqno(2.18)$$

The oscillator model (2.17) yields wave functions that are Gaussian in the
radius $R$, with the lowest three eigenvalues
$$\epsilon_{00}=\sqrt{18}\omega,\quad \epsilon_{01}=\sqrt{24}\omega,
\quad   
\epsilon_{10}=\epsilon_{02}=\sqrt{30}\omega.\eqno(2.19)$$
The ``funnel'' model (2.18) yields wave functions that  are more 
sharply peaked at small 
$R$ and flatter than Gaussians at large $R$. It yields a 
better spectrum than the oscillator model since it splits the SD shell in the
baryon spectrum, which
is degenerate in  the oscillator model.
For the funnel model he dependence of the eigenvalues on the  parameters $a$
and $b$ must be determined numerically.
The ground state wave functions of these two models are shown in
in Fig. 1 for the parameters chosen in Section 4.1.

There is, of course, a wide variety of other possibilities for the
confining function $f(R)$.  A version intermediate  between the
oscillator and the funnel may be obtained by replacing the
linear powers of $R$ in (2.18) by $R^2$ in both terms. That function
 yields  energy levels  between
those of the oscillator and funnel models.

\newpage

\centerline{2.3 The Hyperfine Mass Splitting}
\vspace{0.5cm}

The baryon spectrum in the light flavor sectors is complicated by
the fact that it requires  hyperfine splitting sufficiently strong to
mix up the shell structure of the confining well. The clearest
manifestation of this is the positive parity of the $N (1440)$ and the $\Delta
(1600)$ which are  the  lowest excited states of
the nucleon and the $ \Delta (1232)$ respectively.
A principal  advantage of the study of  heavy flavor hyperons is 
a spectrum  presumably less sensitive to the hyperfine
interaction, and thus easier to interpret.

We consider the following phenomenological model for the hyperfine
interaction:

$${\cal M}^{'} = \{ 1-A [(\vec {r} \times
\vec {k})^{2} + (\vec {\rho} \times \vec{\kappa})^{2}] \}
H_\chi, \eqno(2.20)$$
where $H_\chi$ is the following flavor-spin operator:

$$ H_\chi = - \sum_{i < j}   \biggl\{  \sum_{a=1}^{3} C
\lambda ^{a} _{i} \lambda ^{a} _{j} + 
\sum_{a=4}^{8} C_S \lambda^{a}_{i}
\lambda^{a}_{j} + \sum_{a=9}^{12} C_{C} \lambda ^{a}_{i} 
\lambda ^{a} _{j} + 
\sum_{a=13}^{14} C_{SC} \lambda ^{a}_{i} 
\lambda ^{a} _{j} $$
$$+ \sum_{a=16}^{19} C_{B} \lambda^{a} _{i}
\lambda^{b}_{j}+ \sum_{a=20}^{21} C_{SB} \lambda^{a} _{i}
\lambda^{b}_{j} \biggr\}
\vec \sigma_i\cdot\vec \sigma_j .\eqno (2.21)$$
The matrices  $ \lambda ^{a}_{i} $ 
are  $SU(5)$ representations -
extensions of the $SU(3)$ Gell-Mann matrices. 
The parameters
$C,C_S,C_C, C_{SC}, C_B$ and $C_{SB}$ must  
be determined by
the empirical spectrum. The angular momentum
dependent term in eq. $(2.13)$ is motivated by the substantial
empirical splitting of the SD-shell in the light flavor sector (cf. the
splitting between the $N(1440)$ and 
the $ N(1720)-N(1680)$ multiplet).

Since the contributions ${\cal M}_0$ and ${\cal M}'$
commute, the eigenvalues of the combined mass operator
${\cal M}_0+{\cal M}'$ are additive, that is
$$
{\cal E}={\cal E}_0+c\; ,
\eqno(2.22)$$
where $c$ is an eigenvalue of ${\cal M}'$. The hyperfine
correction $c$ depends on the orbital angular momentum
of the 3 quark state and on the symmetry 
character of the spin-flavor part of the wave function.

If the hyperfine term (2.20)
is viewed as an effective representation of pseudoscalar
exchange mechanisms between quarks, the matrix elements $C$
represent $\pi$, and the matrix elements $C_S$, $K$ and $\eta$ meson
exchange. The matrix elements $C_C$ and $C_{SC}$ would
represent $D$ and $D_s$ exchange mechanisms respectively
and correspondingly the matrix elements $C_B$ and $C_{SB}$
would represent $B$ and $B_s$ exchange mechanisms. 
Because of the near degeneracy
between the heavy flavor pseudoscalar and vector mesons, the
matrix elements $C_S$ and $C_B$ should
be viewed as representing
both heavy flavor pseudoscalar and vector exchange
mechanisms. The terms $a=15$ and $a=22$ have been left out from
the sums in the hyperfine interaction (2.20), because they
would represent hidden heavy flavor $\eta_C, J/\psi$ and 
$\eta_B, \Upsilon$ 
exchange mechanisms that
should be far weaker than the corresponding heavy flavor
pseudoscalar and vector exchange mechanisms, because of
the much larger  $c\bar c$ and $b\bar b$ masses.

\newpage

\centerline{\bf 3. The Current Operator}

\vspace{1cm}
\centerline{3.1 Covariance Conditions}
\vspace{0.5cm}

The quark current density operators $I^\mu(x)$ have to satisfy
the covariance conditions
$$U^\dagger(\Lambda)I^\mu(x)
U(\Lambda)=\Lambda^\mu\, _\nu I^\nu(\Lambda^{-1}x)\; ,
\eqno(3.1)$$
for arbitrary Lorentz transformations $\Lambda$, and
$$e^{iP\cdot a}I^\mu(x)e^{-P\cdot a}=I^\mu(x+a)\; .
\label{TRC}\eqno(3.2)$$
for space time translations. Current conservation requires that
$$[P_\nu,I^\nu(0)]=0\; .\eqno(3.3)$$
Let $|M,\vec P,j,\sigma,\tau,\zeta>$  be
eigenstates of the four momentum operator
$P=\{\sqrt{\vec P^2+M^2},\vec P\}$
and the canonical spin,
where $\sigma $ is an 
eigenvalue of $j_z$ and $\zeta=\pm 1$ is the
intrinsic parity. Lorentz invariant form factors are matrix elements of
the form
$$<\kappa',\tau',\sigma',j',\vec P\,',M'|
{I}^\mu(0) |M,\vec P,j,\sigma,\tau,\kappa>,
\eqno(3.4)$$
Because of the covariance of the current operator and the
basis states only matrix elements with 
$\vec P\,'=-\vec P\equiv \vec Q/2$
are required, and thus
the initial and final states are related kinematically.
Note that $\vec Q\,^2$:
$$\vec Q\,^2 = Q^2+\frac{(M'^2-M^2)^2}{Q^2+2(M'^2+M^2)},
\eqno(3.5) $$
is Lorentz invariant. We may assume, without loss of generality, 
that the $z$-axis 
is in the direction
of $\vec Q$. 

The current conservation condition (3.3) is then expressed in the form
$$
\left[1+{4{\cal M}^2\over\sqrt{\vec Q\,^2 }},\rho (0)\right] - I_L(0)
=0.\eqno(3.6) $$
where the Lorentz invariant charge and longitudinal current density operators 
$\rho(0)$ and $I_L(0)$ are defined as
$$
\rho(0):= {1\over 2}\{{\cal M}^{-1}P_\mu, I^\mu(0)\},  \qquad \qquad 
\sqrt{\vec Q\,^2} I_L(0):=Q\cdot I + 
[\sqrt{\vec Q\,^2/4+{\cal M}^2},\rho(0)]
\eqno(3.7) $$
Matrix elements
$<f| I^\mu(0)|a>$ of the current density are then related to the wave functions
and current kernels by
by 
$$<f| I^\mu(0)|i>= \int d^3 p_1\,'\int d^3 p_2\,' 
\int d^3 p_3\,'\int d^3 p_1
\int d^3 p_2 \int d^3 p_3$$
$$ \times
\Psi^*_f(\vec p_1\,',\vec p_2\, ',\vec p_3\, ')
(\vec p_1\,',\vec p_2,',\vec p_3\,'|I^\mu(0)|\vec p_3,\vec p_2, \vec p_1)
\Psi_a(\vec p_1,\vec p_2,\vec p_3)\; .
\eqno(3.8)$$

Elastic electric and magnetic form factors, $G_E(Q^2)$ and $G_M(Q^2)$  
are defined by
$$
G_E(Q^2):= {1\over 2 j+1}\sum_\sigma
\langle j ,\sigma, \vec Q/2|\rho(0)|-\vec Q/2,j,\sigma\rangle\; ,
\eqno(3.9a)$$
$$
G_M(Q^2)=\frac{-1}{\sqrt{\eta}}\sum_{\sigma',\sigma}
(-)^{j-\sigma}(j ,j,\sigma',-\sigma|1,1)
\langle j ,\sigma',\vec Q/2|I_x(0)+iI_y(0)|-\vec Q/2,j,\sigma\rangle\; ,
\eqno(3.9b)
$$
where $\eta:=XCQ^2/4m_p^2$ and $m_p$ is the proton mass.
Magnetic moments, 
expressed in terms of nuclear magnetons,
are equal to the magnetic form factors at $Q^2=0$.

\vspace{1cm}
\centerline{ 3.2  Impulse Currents}

\vspace{0.5cm}

Impulse currents may be defined by kernels of the charge and transverse
currents,
$$(\vec p_1\,',\vec p_2\,',\vec p_3\,'|\rho|\vec p_3,
\vec p_2,\vec p_1)
=(\vec p_1\,'|\rho_1|\vec p_1)\delta^{(3)}(\vec p_2\,'-\vec p_2)
\delta^{(3)}(\vec p_3\,'-\vec p_3)$$
$$+(\vec p_2\,'|\rho_2|\vec p_2)\delta^{(3)}(\vec p_3\,'-\vec p_3)
\delta^{(3)}(\vec p_1\,'-\vec p_1)$$
$$+(\vec p_3\,'|\rho_3|\vec p_3)\delta^{(3)}(\vec p_1\,'-\vec p_1)
\delta^{(3)}(\vec p_2\,'-\vec p_2).
\eqno(3.10)
$$
and
$$(\vec p_1\,',\vec p_2\,',\vec p_3\,'|I_\perp|\vec p_3,
\vec p_2,\vec p_1)
=(\vec p_1\,'|I_{\perp, 1}|\vec p_1)\delta^{(3)}(\vec p_2\,'-\vec p_2)
\delta^{(3)}(\vec p_3\,'-\vec p_3)$$
$$+(\vec p_2\,'|I_{\perp, 2}|\vec p_2)\delta^{(3)}(\vec p_3\,'-\vec p_3)
\delta^{(3)}(\vec p_1\,'-\vec p_1)$$
$$+(\vec p_3\,'|I_{\perp,3}|\vec p_3)\delta^{(3)}(\vec p_1\,'-\vec p_1)
\delta^{(3)}(\vec p_2\,'-\vec p_2).
\eqno(3.11)
$$

Because of the complete antisymmetry of the baryon
wave functions it is sufficient to consider the current
matrix elements of only one constituent, e.g. $i=3$.

The simplest  model for the electromagnetic current density
for the light and strange baryons,  
which   matches the features of 
conventional quark models,
is  specified by kernels $(\vec p_i\,'|\vec {\cal I}|\vec p_i)$ 
that depend
on the momentum transfer $\vec p_i\,'-\vec p_i=\vec Q$, the spin and
flavor
variables, but do not depend on
 $\vec p_i\,'+\vec p_i$. That is
$$(\vec p_i\,'|\rho_i|\vec p_i)=
\left[{1\over 2}
\lambda_3^{(i)}f_3(\vec Q^2)+
{1\over 2\sqrt{3}}\lambda^{(i)}_8f_8(\vec Q^2) \right]\otimes 1
\; .
\eqno(3.12)$$
$$
(\vec p_i\,'|I_{\perp i}|\vec p_i)=i{\vec \sigma_i\times \vec Q\over 2 m_u}
\left\{{2\over 3}+{1\over \sqrt{3}}\lambda_8\right\}
\left[{1\over 2}
\lambda_3^{(i)}g_3(\vec Q^2)+
{1\over 2\sqrt{3}}\lambda^{(i)}_8 g_8(\vec Q^2) \right]
\otimes 1
$$
$$
\hspace*{3cm}+
i{\vec \sigma_i\times \vec Q\over 2 m_s}
\left\{{1\over 3}-{1\over \sqrt{3}}\lambda_8\right\}
\left[{1\over 2}
\lambda_3^{(i)}g_3(\vec Q^2)+
{1\over 2\sqrt{3}}\lambda^{(i)}_8 g_8(\vec Q^2) \right]
\otimes 1\; .
\eqno(3.13)$$
The functions $f_3(\vec Q^2)$, $f_8(\vec Q^2)$ and 
 $g_3(\vec Q^2)$, $g_8(\vec Q^2)$
represent charge and magnetization distributions
of the constituent quark.  The scale factors  $m_u=m_d$ and $m_s$
represent quark masses.

A straightforward extension  to include the
charm and beauty  is achieved by
adding  terms of the form
$${1\over 6}f_C(\vec Q^2)
[1-\sqrt{6}\lambda^{(i)}_{15}]-
{1\over 15}f_B(\vec Q^2)
[1-\sqrt{10}\lambda^{(i)}_{22}] \eqno(3.14)$$
to the quark charge operators.

\vspace{1cm}
\centerline{ 3.3 The Form Factors of the Nucleons.}

\vspace{0.5cm}

The charge density operator  (3.12)  yields  the following
expressions for the electric form factors of the proton and neutron.
$$
G_E^p( Q^2)= {1\over 2} [f_8(\vec Q^2)+f_3(\vec Q^2)]
F_0(\vec Q^2)\; ,
$$
$$
G_E^n( Q^2)= {1\over 2} [f_8(\vec Q^2)-f_3(\vec Q^2)]
F_0(\vec Q^2)\; .
\eqno(3.15)$$
The function $F_0(\vec Q^2)$ is related to the ground-state
wave function $u_00(R)$ by the Fourier-Bessel transform
$$F_0(\vec Q^2):={3\pi\over 16 \tilde Q^2}\int_0^\infty
dR R^3 J_2(\tilde QR)u_{00}^2(R)\; , 
\eqno(3.16)$$
where the variable $\tilde Q$ is defined as
$$
\tilde Q:=\sqrt{\frac{\vec Q^2}{1+\frac{\vec Q^2}{4m_p^2}}}\; .
\eqno(3.17)$$
The normalization of the wave function implies $F_0(0)=1$.
From $G_E^p( 0)=1$ and $G_E^n( 0)=0$ it
follows that $f_3(0)=f_8(0)=1$.

The proton  charge radius  obtains  as
$$
<r^2_p>=-6 \left({d G_E^p( Q^2)\over dQ^2}\right)_{Q^2=0}=
-6 \left({d F_0( Q^2)\over dQ^2}\right)_{Q^2=0}
-{1\over 2}\,\left(6{df_3(Q^2)\over dQ^2}+6{df_8(Q^2)\over dQ^2}\right)_{Q^2=0}
$$
$$
=
\frac{\pi}{768}\int_0^\infty dR R^7 u_{00}^2(R)
+\frac{1}{2}(<r_8^2>+<r_3^2>)\; .
\eqno(3.18)$$
Here $<r_8^2>$ 
and $<r_3^2>$ may  be interpreted as mean square isoscalar and isovector
quark charge radii  respectively. 

Assuming point-quark charges, $f_3=f_8=1$, is manifestly inadequate
as the wave functions for both the oscillator or the funnel model 
yield much too small proton charge radii, and form factors 
that are too large for all values of the  momentum
transfer. This defect is cured by assuming appropriate
quark-charge distributions. 

Observed values of the e charge form factors of proton and neutron can
can be reproduced with the funnel model 
 assuming
$$
f(\vec Q\,^2):={1\over 2}[f_8(\vec Q\,^2)+f_3(\vec Q\,^2)]
=\frac{1}{1+\vec Q\,^2/\Lambda^2}\; , 
\eqno(3.19a)$$
and
$$
{1\over 2}[f_8(\vec Q\,^2)-f_3(\vec Q\,^2)]= 
{\,Q^2 
f(\vec Q\,^2) \over 4 \Lambda^2 \sqrt{1+ 7\vec Q^2/4\Lambda^2}}\; .
\eqno(3.19b)
$$
The value $\Lambda=$ .7 GeV is designed to fit the proton charge radius.
The corresponding quark-charge radius is  
$\sqrt{\langle r_q^2\rangle}\approx .7$ fm.
The slope of $G_E^n(Q^2)$ at $Q^2=0$
$$
{d G_E^n(Q^2)\over dQ^2}={1\over 4 \Lambda^2}= .51 \mbox{GeV}^{-2}
\eqno(3.20)
$$
is equal to the observed value \cite{KroRi,KoeNW}.
The  $Q^2$ dependence of the proton charge form factor is 
illustrated in Fig. 2a in 
comparison with the  data tabulated in ref.\cite{Hoe}.
We also show the charge form factors
calculated with point-quark charge distributions, $f_3=f_8=1$, and with the
wave functions of  the oscillator and funnel models,
The potential parameter  are determined by the mass spectra in Sec. 4
below. Fig. 2b illustrates the $Q^2$ dependence of the neutron charge
form factor
together with and  empirical fit \cite{Hoe}, which is consistent with more 
recent data\cite{Plat}.

From the current density operator  (3.13) it follows that  the magnetic
form factors of the nucleons depend on the the quark structure
and the wave functions according to
$$G_M^p(\vec Q^2)= {m_p\over 6m_u} [g_8(\vec Q^2)+5g_3(\vec Q^2)]
F_0(\vec Q^2)\; ,
$$
$$
G_M^n(\vec Q^2)= {m_p\over 6m_u} [g_8(\vec Q^2)-5g_3(\vec Q^2)]
F_0(\vec Q^2)\; .
\eqno(3.21)$$
Since the nucleon magnetic form factors can be approximated by the proton
electric form factor multiplied with the magnetic moments
\cite{Hoe} we choose $m_u=.336$ GeV to fit the magnetic moment of the proton
with $g_8(0)+5g_3(0)=1$. Magnetic form factors with
the correct order of magnitude 
obtain with $g_8(\vec Q^2)=g_3(\vec Q^2)=f(\vec Q\,^2)$.
A reasonable representation of the magnetic form
factors of the nucleons is achieved by
$$
g_8(\vec Q^2)={\mu_8\over 1+\vec Q^2/1.3\Lambda^2}
\; ,\qquad g_3(\vec Q^2)={\mu_3\over 1+\vec Q^2/1.3 \Lambda^2}
\eqno(3.22)
$$
with $\mu_8=.95$ and $\mu_3=1.01$.
In Fig.3 we show the  magnetic
form factors of the funnel  model  in comparison with the
empirical fit of  ref. \cite{Hoe}. 

\vspace{1cm}

\centerline{ 3.4  Magnetic Moments 
and Axial Coupling Constants of the Ground State Baryons }

\vspace{0.5cm}

The magnetic moments that are obtained from the current
operator (3.13) are the usual linear
combinations of proton to quark mass ratios listed in
the Table 1 along with the corresponding
empirical values \cite{PDG,Boss} for the nucleons and
the strange hyperons. The strange scale factor $m_s$
is determined by the
 magnetic moments of the
$\Omega^-$ to be  $m_s=$ .465 GeV. 
The corresponding expressions for the magnetic moments of
the heavy flavor baryons are given eq. in refs.
\cite{GloRis1, Lich}.

A model for the axial current density, which corresponds to that
of the vector current (3.13) is
$$\vec {\cal A}_i(\vec Q)=g_A^q {\lambda_a^{(i)}\over 2}
\vec\sigma_i\; .
\label{AXS}
\eqno (3.23)$$
Here $g_A^q$ represents the axial coupling constant for a
single constituent quark. This model for the axial current
leads to the same results  as the conventional quark model
for the $g_A/g_V$ ratios of the nucleons and the strange
hyperons (Table 2). 
In the conventional representation of the $g_A/g_V$
ratios for the hyperon decays in terms of $D$ and
$F$ coefficients, these take their usual quark model
values
$$D=1,\quad F={2\over 3}\; .
\eqno (3.24)$$
Agreement of this model  with the empirical value for
the axial vector coupling for neutron decay  requires
the value $g_A^q=0.76$. 

\vspace{1cm}

\centerline{\bf 4. The Spectra of the Nucleon and the $\Delta$
resonances}
\vspace{0.5cm}

	The spectra of the
nucleon and the $\Delta$ resonances depend on the choice of
the parameters $C$, $C_S$ and $A$ in the hyperfine interaction
(2.20). The expressions for these hyperfine shifts of the states
in the $S$, $P$, and $SD$ shells of these spectra are listed
in Table 3, where we have employed the symmetry assignments of
ref. \cite{GloRis}. The real values of the empirical pole
positions of the known resonances given in ref. \cite{PDG} have 
also been listed in Table 3.

	With the parameter values $a=5.66$ fm$^{-1}$
and $b=6.09$ fm$^{-3}$ in the mass operator model (2.18)
and with the parameter values
$C=$ 28 MeV, $C_S=$ 19 MeV and $A=$ 0.22 
in the hyperfine term (2.21) we obtain 
very satisfactory energies for all the known states.
The eigenvalues $\epsilon_{nK}$ determined by the first
two of these values are
$$\epsilon_{00}=1340 {\rm MeV},\epsilon_{01}=1652 {\rm MeV},
\epsilon_{10}= 1777 {\rm MeV}, \epsilon_{02}=1867 {\rm MeV}.
\eqno (4.1)$$
The same value for $\epsilon_{00}$ obtains for the
oscillator model (2.17) with $\omega=$ 316 MeV.
The ground state wave functions plotted in Fig. 1 have
been calculated with these parameter values.

The value for $C_S$ is directly determined by the empirical
mass splitting between the $\Sigma$ and  $\Sigma(1385)$
hyperons (cf. Table 5 and section 5.1 below):
$$m[\Sigma(1385)]-m[\Sigma]=10 C_S, \eqno(4.2)$$
The value for the parameter $C$ is then chosen so as to obtain
agreement with the empirical $N\Delta$ splitting (Table 3):
$$m[\Delta(1232)]-m[N]= 12 C - 2 C_S.\eqno(4.3)$$
Finally the value for the parameter $A$ is  chosen
so as to obtain agreement with the
splitting between the $SD$ shell
multiplets $N(1440)$ and $N(1720)-N(1680)$ multiplets.

	The calculated resonance energies in Table 3 agree
with the empirical pole positions
of the known states to within 2 \% in most cases. 
A number of the still missing states have been 
identified in recent phase shift analyses. The ${3\over 2}^+$
member of the $L=2$\,\, $\Delta$ multiplet near 1750 MeV has
been located in ref. \cite {Man} at 1754 MeV. In that
analysis a $P_{13}$ state has also been located at 1879 MeV: this
corresponds well to the $L=0$ nucleon resonance 
predicted at 1881 MeV. The additional 
$P_{11}$
resonance predicted at 1916 MeV may be related to the 
additional, if somewhat lower lying $P_{11}$ resonance
found in the phase shift analysis \cite{Nef}.  
\newpage


\centerline{\bf 5. The Strange Hyperons}
\vspace{0.5cm}
\centerline{5.1 The $\Lambda$ and $\Sigma$
spectra}
\vspace{0.5cm}

The parameter values above may be employed directly to the 
spectrum of the strange $S=-1$ hyperons. The only additional parameter
required is the mass parameter $\Delta_S$ in (2.9). With 
$\Delta_{S}=587$ MeV and the confining
potential (2.18) the spectrum of the $\Lambda$ -hyperon
is obtained satisfactorily as shown in Table 4.
The one missing feature is the large spin-orbit splitting of the
$\Lambda(1405)-\Lambda (1520) $ multiplet, which is expected to have
an anomalously large $\bar{K} N $ -admixture \cite {Dal}. 
The peculiar aspect of this flavor singlet multiplet is brought
out by the fact
that it is the only $P$-shell baryon multiplet for which the
empirical spin-orbit splitting is not consistent with zero.

The quality of the $\Lambda$ spectrum is similar to that
of ref. \cite{GloRis}. 
The good agreement
between the calculated and the empirical spectra lends credence to the
quantum number and symmetry assignments made in Table 4.
Note that one $3/2^-$ state 
in the $P$ shell of the $\Lambda$ spectrum near 1800 MeV
remains to be found experimentally. This state would
correspond to the $N(1700)$ resonance in the nucleon spectrum.

The still uncertain quantum number assignments in the 
empirical $\Sigma$ hyperon spectrum rule out a definite
assessment of the quality of the predicted $\Sigma$ spectrum
shown in Table 5, which was obtained with the same parameter
values as the $\Lambda$ spectrum in Table 4.
The energies of low lying well established positive parity
$\Sigma$ resonances $\Sigma(1385)$ and $\Sigma(1660)$ and the
negative parity $\Sigma(1775)$ and $\Sigma(1915)$
resonances are however in satisfactory agreement
with the empirical energies. 
        
The calculated spectrum in Table 5 suggests that the
$\Sigma(1840)$ be the $S=-1$ analog of the $\Delta(1600)$
breathing resonance. The absence of low lying $\Sigma$
resonances in the predicted spectrum also suggests
that 
the unassigned and unconfirmed $\Sigma(1480)$ and
$\Sigma(1560)$ states may turn out to be spurious. The experimental
information on the $P$ shell of the $\Sigma$ spectrum
is still incomplete: both a $1/2^-$ and a $3/2^-$ state
are still missing.

The interpretation of the $\Sigma$ hyperon spectrum is
complicated by the appearance of 2 negative parity
resonances - the $\Sigma(1940)$ and the $\Sigma(2000)$ -
that belong to the $PF$ shell in the vicinity of
2 GeV. In the spectrum of the $\Lambda$ the corresponding
$PF$ shell resonances lie at higher energies (e.g. the
$\Lambda(2100), 7/2^-$ resonance). In the nucleon spectrum
the $PF$ shell also begins at correspondingly higher
excitation (eg. the $N(2080), 3/2^-$ resonance).

\vspace{1cm}
\centerline{5.2 The $\Xi$ spectrum}
\vspace{0.5cm}

The spectrum of the $S=-2$ $\Xi$ hyperon
is given in Table 6. In the absence of empirical
assignments for the excited $\Xi$ states we have only
indicated those of the well established $\Xi$, $\Xi(1530)$
and $\Xi(1820)$ states. The energies of these states are
in good agreement with the corresponding
empirical values. The recent discovery of a state at
1690 MeV \cite{WA89} agrees well with the predicted
$1/2^+$ state at 1695 MeV. This state is the $S=-2$ 
analog of the $N(1440)$, $\Lambda(1600)$ and the 
$\Sigma(1660)$. 

The presently known empirical $\Xi$ spectrum contains
candidates for many  states of the present model in Table 6. 
The calculated energies are close to those predicted in
ref. \cite {GloRis}.

\vspace{1cm}

\centerline{5.3 The $\Omega^-$ spectrum}

\vspace{0.5cm}

In the present model the structure of the the
$\Omega^-$ spectrum is similar to that of the $\Delta$
resonance. The spectrum in Table 7 has 3
positive parity multiplets with energies in the region
2.1 -- 2.2 GeV. The empirical $\Omega^-(2250)$
very likely corresponds to one of these. The empirical
$\Omega^-(2380)$ may also be a positive parity
resonance in the $SD$-shell, but the $\Omega^-(2470)$ 
resonance is probably
a negative parity resonance in the $PF$ shell.

As the mass of the $\Omega^-$ differs by 66 MeV
(i.e. by $\simeq$ 4\%) from its experimental value
that amount should be added to
the energies of the $\Omega^-$ resonances.
With that shift all the excited states of the
$\Omega^-$ lie above 2 GeV. This also
suggests the two highest $D$-shell resonances 
in Table 7 may be interpreted as the
$\Omega(2380)^-$ and the $\Omega(2250)^-$
respectively. 

With the model (2.20)-(2.21) for the hyperfine interaction
the baryon mass spectra
 satisfy the same
 relations as  the model in ref. \cite{GloRis}, e.g.
the relations
$$m[\Delta(1232)]-m[N]=m[\Sigma(1385)]+{3\over 2}(
m[\Sigma]-m[\Lambda]),$$
$$m[\Sigma(1385)]-m[\Sigma]=m[\Xi(1530)]-m[\Xi],$$
$${1\over 3}(m[\Omega]-m[\Delta])=m[\Xi(1530)]-m[\Sigma(1385)].
\eqno(5.1)$$
These relations are in good agreement with the experimental values.
The Gell-Mann-Okubo and equal spacing rules are recovered
in the $SU(3)_F$ symmetric limit $C=C_S$.

\vspace{1cm}
\centerline{\bf 6. The Charm Hyperons}

\vspace{0.5cm}

\centerline{ 6.1 The C=1 Hyperons}

\vspace{0.5cm}

\centerline{6.1.1 The $\Lambda_C^+$ and $\Sigma_C$ Spectra}

\vspace{0.5cm}

The hitherto observed charm hyperons all have $C=1$, i.e. they contain
one charm  quark and two light quarks. While the charm
hyperon spectrum in many aspects is expected to be similar to
the strange hyperon spectrum, it differs from that in the much
larger number of narrow states, which lie below the threshold for
strong decay. In the spectrum of the $\Lambda$ there is only
one state below the threshold for $\bar K N$ decay - the
$\Lambda(1405)$. In contrast all the 7 negative parity states
in the P-shell as well as the positive parity breathing resonance of the
$\Lambda_C$ are expected to lie well below the threshold for
$\bar D N$ decay. This relative compression  of the charm hyperon spectra to
lower excitation energies is achieved  through the 
flavor dependence, ${\cal K}$ of the mass operator ${\cal M}_0$.
A larger contribution  ${\cal K}$
reduces the repulsive effect of the momentum dependent
terms in ${\cal M}_0$.

The spectra of the non-strange $C=1$ hyperons depend on
two additional parameters:
the charm quark parameter $\Delta_C$ in (2.9) and the parameter
$C_C$ in the hyperfine interaction (2.20). The expressions for the
energies of the states in the spectra of the $\Lambda_C^+$ and
$\Sigma_C$ are listed in Tables 8 and 9 respectively.

The hyperfine interaction parameter $C_C$ - which in ref. 
\cite {GloRis1} was interpreted as the matrix element of
$D$ and $D^*$ meson exchange interactions - is determined
by the mass difference between the $\Sigma_C$ and the
$\Sigma_C(1530)$ as (cf. Table 9):
$$m[\Sigma_C(2530)]-m[\Sigma_C]=6 C_C.\eqno(6.1)$$
The new  value 56 MeV $\pm$ 3 MeV \cite{StaBar} for the mass splitting
between the 
$\Sigma_C(2530)$ and the $\Sigma_C$ then yields $C_C\simeq$ 10 MeV.  
With this value
for $C_C$ the expression for the mass of the $\Lambda_C^+$
in Table 8 implies that the empirical value for that mass
is reached with $\Delta_C=$ 2.2 GeV. We  use this
value here.  

Given these parameter values the masses of all the other
states in the spectra of the $\Lambda_C^+$ and the $\Sigma_C$
are determined. We note the agreement of the model with the
average mass value of of the negative parity multiplet
$\Lambda_C(2593)^+ - \Lambda_C(2625)^+$. This multiplet is
the charm analog of the $\Lambda(1405)-\Lambda(1520)$
multiplet. The satisfactory simultaneous description of
these two multiplets by the present model supports view
that the former is a flavor singlet combination of
$u$, $d$ and $c$ quarks, and the latter is a 
flavor singlet combinations of $u$, $d$ and $s$ quarks. 

At the present time no excited states of the $\Sigma_C$ beyond the
$\Sigma_C(2530)$ have been found. The present prediction is that
the $P-$shell states of the $\Sigma_C$ fall within the range
2.6 -- 2.7 GeV. This is slightly higher than the corresponding
predictions in \cite {GloRis1}. The positive parity breathing 
resonances of both the $\Sigma_C$ and the $\Sigma_C(1530)$ are predicted 
to coincide with the lowest $P$-shell states.

\vspace{1cm}
\centerline{6.1.2 The $\Xi_C$ Spectrum}

\vspace{0.5cm}

The hyperons with $C=1$ and $S=-1$ are denoted $\Xi_C$. In the
constituent quark model these states are formed of one light,
one strange and one charm quark. It is natural to divide the
spectrum of the $\Xi_C$ into two separate sectors,
which are respectively antisymmetric (denoted $\Xi_C^a$)
and symmetric (denoted $\Xi_C^s$) under interchange of
the light and the strange quark. This is in 
analogy with the separation of the $S=-1$ hyperon spectra
into separate $\Lambda$ and $\Sigma$ sectors. The structure
of the $\Xi_C^a$ spectrum should thus be similar to that of the
$\Lambda$ and that of the $\Xi_C^s$ spectrum similar to that
of the $\Sigma_C$ hyperon.

	Both the $\Xi_C^a$ and the $\Xi_C^s$ spectra  
differ from those of the $\Lambda$ and the
$\Sigma$ hyperons in that their entire $P$-shells
are expected to lie below the threshold for the strong decay to
$\bar D N$ (2907 MeV). Their low lying negative parity
states should therefore be sharp. This expectation appears
to be confirmed by the
narrow width ($<$ 2.4 MeV) of the recently found ${3\over 2}^-$
$\Xi_C^a$ resonance at 349 MeV excitation \cite{CLEO}.

The expressions for the energies of the predicted $S$,
$P$ and $SD$ shell states of the $\Xi_C$ hyperons are listed in
Tables 10 and 11. These energies depend on one 
new parameter: $C_{SC}$,  determined by
the splitting between the $\Xi_C^s$ and the $\Xi_C^a$ as
$$m[\Xi_C^s]-m[\Xi_C^a] = {20\over 3}C_S - 2C_C-2C_{SC}.
\eqno(6.2) $$ 
With the present  uncertain value of the $\Xi_C^s$ (2560 MeV)  mass
 we have $C_{SC}$ = 5 MeV. With
this  value of $C_{SC}$  and those fixed previously
 we obtain the
complete $S$, $P$ and
$SD$ shell spectra of the $\Xi_C^a$ and  $\Xi_C^s$. 
In ref. \cite {GloRis1} was
interpreted as the matrix element of $D_s$ and $D_s^*$
exchange interactions between the $s$ and $c$ quarks.

The ${3\over 2}^-$ state found in ref. \cite {CLEO} at an
excitation energy of 349.4 $\pm$ 0.7 $\pm$ 1.0 MeV
above the $\Xi_C^a$ was interpreted as the strange-charm
analog of the $\Lambda_C(2625)^+$ ${3\over 2}^-$ state.
As in the present model the corresponding state is predicted
to have the considerably lower excitation energy 269 MeV,
the ${3\over 2}^-$ state member of the spin 3/2 negative
parity multiplet is predicted to have an excitation energy
of 378 MeV, we suggest that this is the more
natural assignment for the ${3\over 2}^-$ state at 349 MeV
excitation energy, than the $usc$ flavor singlet state.

\vspace{1cm}
\centerline{6.1.3 The $\Omega_C^0$ Spectrum}

\vspace{0.5cm}

The spectrum of the $\Omega_C^0$ is the only sector of the
$C=1$ hyperon spectrum, in which the ${3\over 2}^+$ state
in the $S$-shell has yet to be found experimentally. This
is expected to lie about 30 MeV above the ground state
$\Omega_C^0$. In the present model the mass of the 
$\Omega_C^0$ is 77 MeV -- i.e. by $\sim$ 3\% -- smaller
than the experimental value.
This difference is of similar magnitude as that for the
$\Omega^-$ (66 MeV), and could of course be eliminated by
slightly 
increasing the factor multiplying the parameter $\Delta_S$
in the mass operator (2.9). While this mass difference
is relatively small, it may indicate that better
energies for the excited states
of the $\Omega_C^0$ would obtain if an amount 77 MeV were
added to the calculated energies in Table 12, where
the expressions for all the states in the $S$, $P$ and
$SD$ shells of the $\Omega_C^0$ are listed.

\vspace{1cm}
\centerline{6.2 The $\Xi_{CC}$ and $\Omega_{CC}$ Spectra}

\vspace{0.5cm}

The hyperons with $C=2$ are denoted $\Xi_{CC}$. All
members of this class of particles remain to be discovered.
As the hyperfine interaction between the 2 charm quarks
in these hyperons is expected to be of negligible
magnitude compared to that between the light quark and the
charm quarks, their spectrum should be easier
to interpret than that of the lighter hyperons.

The present model for the hyperfine interaction
(2.20), (2.19) implies that the structure of the spectrum
of the $\Xi_{CC}$ hyperons should be similar to that of
the $\Sigma_C$ hyperons. In the present model the hyperfine 
splittings of the $\Xi_{CC}$ spectrum are completely
determined by the parameters $C$ and $A$ (2.20),(2.21).
The masses of the $\Xi_{CC}$ ($1/2^+$) and the $\Xi_{CC}^*$
($3/2^+$), which form the $S$ shell of the $\Xi_{CC}$
spectrum, are given as
$$m[\Xi_{CC}]=\sqrt{\epsilon_{01}^2+2\Delta_C^2}-10C=3290
{\rm MeV},\eqno(6.3)$$  
$$m[\Xi_{CC}^*]=\sqrt{\epsilon_{01}^2+2\Delta_C^2}-4C=3350
{\rm MeV}. \eqno(6.4)$$  
While these absolute mass values are likely to be too low
by 100 -- 200 MeV, the splitting (60 MeV) 
between
the $\Xi_{CC}$ and the $\Xi_{CC}^*$ is probably a realistic
prediction.
The bound state version of the Skyrme model, which typically
gives realistic values for the ground state hyperons \cite
{Rho}, yields a mass of 3.5 GeV for the $\Xi_{CC}$ 
and value for the mass difference between the $\Xi_{CC}$
and the $\Xi_{CC}^*$ in agreement with the present one.

The baryons with $C=2$ and $S=-1$ are denoted $\Omega_{CC}^+$.
The spectrum of these particles can be determined directly
from that of the $\Xi$ hyperons in Table 6. The only
modification is that the constant $C_S$ in the hyperfine
correction has to be replaced by the corresponding
constant $C_{SC}$, which is appropriate for 
simultaneous charm and
strangeness exchange.
The masses of the $\Omega_{CC}^+$ ($1/2^+$) and the 
$\Omega_{CC}^{+*}$
($3/2^+$), which form the $S$ shell of the $\Omega_{CC}^+$
spectrum, are given as
$$m[\Xi_{CC}^+]=\sqrt{\epsilon_{01}^2+2\Delta_C^2
+\Delta_S^2}-10C=3390
{\rm MeV},\eqno(6.5)$$  
$$m[\Xi_{CC}^{+*}]=\sqrt{\epsilon_{01}^2+2\Delta_C^2
+\Delta_S^2}-4C=3420 
{\rm MeV}. \eqno(6.6)$$  
As in the case of the $\Xi_{CC}$ we expect the absolute values 
to be too low by  100 -- 200 MeV, but the splitting
of 30 MeV between the $\Omega_{CC}^+$ and the
$\Omega_{CC}^{+*}$ should be realistic. The Skyrme model
prediction for these two states is  in the region
3.7-3.8 MeV \cite {Rho}.

\vspace{1cm}
\centerline{\bf 7. The Beauty Hyperons}

\vspace{0.5cm}

The present model implies that the spectrum of the $B=-1$ hyperons
be organized very much like that of the $C=1$ hyperons described
above in Tables 8 and 9. The expressions for the energies of
the $\Lambda_B^0$ and the $\Sigma_B$ states are listed in Tables
13 and 14 respectively. These expressions are obtained   
replacing the hyperfine  parameter $C_C$ in the energy
expressions of Tables 8 and 9 throughout by the corresponding 
beauty quark  parameter $C_B$. 

The value of the  parameter $C_B$ is determined
by the empirical splitting between the $\Sigma_B$ and the
$\Sigma_B(5870)$ ):
$$m[\Sigma_B(5870)]-m[\Sigma_B]=6 C_B.\eqno(7.1)$$
With $m(\Sigma_B)=$ 5814 MeV the value of $C_B$ is 
9 MeV. The beauty quark parameter $\Delta_B$ in the mass
operator ${\cal M}_0$ then has to have the value 5.774 GeV in order
that the $\Lambda_B^0$ take its empirical mass 5641 MeV.
Given these parameter values the energies of all the excited
states in the spectra of the $\Lambda_B$ and the $\Sigma_C$
are determined. These energy values are listed in Tables
13 and 14. 

Heavy quark symmetry predicts that the spin of the heavy
quarks decouple from the light quarks, and that therefore
the baryon spectrum should fall into near degenerate 
multiplets with spin $S_l\pm{1\over 2}$, where $S_l$ 
is the spin of the light quarks. Heavy quark
symmetry is implemented in the present model if the
coefficients $C_j$ in the hyperfine interaction (2.21),
which describe interactions between light and heavy
quarks, vanish as the mass of the heavy quarks become
infinite. The fact that the fact that the empirical splittings 
$m[\Sigma_C^*]-m[\Sigma_C]$ and
$m[\Sigma_B^*]-m[\Sigma_B]$ are very similar 
indicates that the limit of heavy quark symmetry 
has not yet been reached in the charm and beauty sectors
of the spectrum.

According to the model outlined here the spectrum of the
$B=-1,S=-1$ $\Xi_B$ hyperons should be organized in the same manner as  the
corresponding spectrum of the $\Xi_C$ hyperons in
Tables 10-11. The energies of the states in the $\Xi_B$
spectrum are obtained from the expressions in
Tables 10-11,replacing the parameters $\Delta_C$, $C_C$
and $C_{SC}$  by the corresponding parameters
$\Delta_B$, $C_B$ and $C_{SB}$. The determination of the
 parameter  $C_{SB}$ from the mass difference
$$m[\Xi_B^s]-m[\Xi_B^a] = {20\over 3}C_S - 2C_B-2C_{BS}.
\eqno(7.2) $$ 
must  await the discovery of the hyperons $\Xi_B^s$ and $\Xi_B^s$.

In our model the energy of the lowest $\Xi_B$ state is 
$$m[\Xi_B^a]=\sqrt{\epsilon_{00}^2
+\Delta_S^2+\Delta_B^2}-{8\over 3}C_S-2C_B-2C_{SB}.\eqno(7.4)$$
If the value of the parameter $C_{SB}$
falls in the range 0 -- 5 MeV, expected from heavy quark symmetry
the mass of
the $\Xi_B^a$  should fall in the
range 5916 -- 5926 MeV. This is larger by 40 -- 50 MeV
than the Skyrme model prediction \cite{Rho}.

\vspace{1cm}

\centerline{\bf 8. Summary}
\vspace{0.5cm}

The present formulation of Poincar\'e covariant quark models
illustrates the possibility of satisfying all the requirements
of relativistic covariance, while reproducing the
good phenomenological results of conventional
constituent quark models for the magnetic moments and
axial couplings along with a good description of the
known baryon spectrum.
The
emphasis in the present work is consistency with general
principles and simplicity while discarding 
the notion of constituent quarks as a system of free
particles that  incidentally are confined in an infinite
potential well.
The key to the satisfactory description of  
the spectrum is a mass operator which is the sum of two commuting terms:
(i) a confining mass operator with the basic point spectrum and (ii) 
a simple spin- and flavor dependent hyperfine term large enough to
affect the level order \cite{GloRis,GloRis1}.

Observable form factors and transition amplitudes 
depend on two distinct features of a quark model that are not separately
observable: (i) the wave function, which  differs for different baryons 
and (ii) the quark-current structure that is the same for all baryons.
Her we have constructed
simple quark-current structures, independent of internal
quark velocity, with instant form kinematics. These models satisfy all
requirements of Poincar\'e covariance  We have shown that
they are consistent with observed nucleon properties.
Electromagnetic form factors
and transition amplitudes of the baryons for all the states
in the $S,\, P$ and $D$ shells of the baryons are well defined.
A quantitative examination of these amplitudes is left to future
investigations. Such investigations may indicate for
configuration mixing in the wave functions.

\vspace{1cm}
{\bf Acknowledgments}

\vspace{0.5cm}

The authors are indebted for instructive discussions on
this topic to L. Ya. Glozman. This
work was supported in part by the U.S. Department of Energy,
Nuclear Physics Division, under contracts W-31-109-ENG-38
and by the Academy of Finland under constract 34081. K.D.
thanks the Arvid and Greta Olin Foundation for financial
support.
\newpage

\newpage

\centerline{\bf Table 1}
\vspace{0.5cm}

Magnetic moments of the ground state baryon octet and the
$\Delta^{++}$ and $\Omega^-$ 
(in nuclear magnetons). Column IA contains the 
impulse 
approximation 
expressions for the current model (3.13). 
Column I contains the numerical values obtained from these 
expressions. The empirical values (``exp'') are from refs.
\cite {PDG, Boss}. 

\begin{center}
\begin{tabular}{|l|l|r|r|} \hline
 & IA & exp & I \\ \hline
 &&& \\
$p$ & ${m_N\over m_u}$ & +2.79  & +2.79 \\
 &&& \\
$n$ & $-{2\over 3}{m_N\over m_u}$ & --1.91 & --1.86\\
 &&& \\
$\Lambda$ & $-{1\over 3}{m_N\over m_s}$ & --0.61 & --0.67 \\
 &&& \\
$\Sigma^+$ & ${8\over 9}{m_N\over m_u}+{1\over 9}{m_N\over m_s}$ &
+2.46 & +2.70\\
 &&& \\
$\Sigma^0$ & ${2\over 9}{m_N\over m_u}+{1\over 9}{m_N\over m_s}$ &
? & +0.84\\
 &&& \\
$\Sigma^0 \rightarrow \Lambda$ & $-{1\over \sqrt{3}}{m_N\over m_u}$
& $|1.61|$ & --1.61  \\
 &&& \\
$\Sigma^-$ & $-{4\over 9}{m_N\over m_u}+{1\over 9}{m_N\over m_s}$
& --1.16 & -1.07  \\
 &&& \\
$\Xi^0$ & $-{2\over 9}{m_N\over m_u}-{4\over 9}{m_N\over m_s}$ &
--1.25 &  --1.52 \\
 &&& \\
$\Xi^-$ & ${1\over 9}{m_N\over m_u}-{4\over 9}{m_N\over m_s}$ &
--0.65 &  --0.59  \\ 
 &&& \\ \hline
 &&& \\
$\Delta^{++}$ & $2{m_N\over m_u}$ & +4.52 & 5.58 \\
 &&& \\
$\Delta^+\rightarrow p$ & ${2\sqrt{2}\over 3}{m_N\over m_u}$ & +3.1 &
 +2.63\\
 &&& \\
$\Omega^-$ & $-{m_N\over m_s}$ & --2.019  & --2.019  \\
 &&& \\
\hline
\end{tabular}
\end{center}
\newpage
\centerline{\bf Table 2}
\vspace{0.5cm}

The axial coupling constants of the baryon octet. Column I gives the
expressions in terms of the F and D coefficients and column II the
predictions for the current operator (3.23)
(with $g_A^q=0.76$). The empirical values are from
refs. \cite{PDG,GaSa}. 

\begin{center}
\begin{tabular}{|l|l|r|r|} \hline
 & I & exp & II  \\ \hline
 &&& \\
$n\rightarrow p$ & $F+D$ & 1.26  & 1.26 \\
 &&& \\
$\Sigma^{\pm}\rightarrow \Lambda$ & $\sqrt{{2\over 3}}D$ & 0.62 & 0.62
\\
 &&& \\
$\Sigma^- \rightarrow \Sigma^0$ & $\sqrt{2}F$ & 0.67 & 0.71 \\
 &&& \\
$\Lambda \rightarrow p$ & $-\sqrt{3\over 2}(F+{D\over 3})$ & 0.92 &
0.96 \\
 &&& \\
$\Sigma^-\rightarrow n$ & $-(F-D)$ & 0.34 & 0.25  \\
 &&& \\
$\Xi^-\rightarrow \Lambda$ & $\sqrt{{3\over 2}}(F-{D\over 3})$ & 0.31
& 0.30 \\
 &&& \\
$\Xi^-\rightarrow \Sigma^0$ & ${1\over \sqrt{2}} (F+D)$ & 1.36 & 0.90 \\
 &&& \\
$\Xi^0\rightarrow \Sigma^+$ & $F+D$ & ? & 1.26 \\
 &&& \\
$\Xi^- \rightarrow \Xi^0$ & $F-D$ & --0.28 & --0.25 \\
 &&& \\ \hline
\end{tabular}
\end{center}
\newpage

\centerline{\bf Table 3}

The nucleon and $\Delta$-states in the $S$, $P$ and $SD$ shells,
The column
$\epsilon$ contains the eigenvalues of the mass operator. The
average over the multiplet of the real part of the empirically
extracted resonance pole position is denoted EXP. The model values
obtained with the mass operators (2.7) with the confining
well (2.18) and the hyperfine interaction (2.20) are
listed below the empirical ones.

\begin{center}
\begin{tabular}{l|l|l|l} \hline
$nKL[f]_{FS}[f]_F[f]_S$ & $LS$ Multiplet & EXP & $\epsilon$
\\ 
 & & (model value) & \\ \hline
$000[3]_{FS}[21]_F[21]_S$ & ${1\over 2}^+, N$ & 939 &
$\epsilon_{00}-15C+C_S$\\
 & & (939) & \\
$000[3]_{FS}[3]_F[3]_S$ & ${3\over 2}^+,\Delta$ & 1209-1211 &
$\epsilon_{00}-3C-C_S$\\
 & & (1237) & \\
$100[3]_{FS}[21]_F[21]_S$ & ${1\over 2}^+,N(1440)$ & 1346-1385 &
$\epsilon_{10}-15C+C_S$\\
 & & (1376) & \\
$011[21]_{FS}[21]_F[21]_S$ & ${1\over 2}^-;N(1535),{3\over
2}^-N(1520)$ & 1496-1527 & $\epsilon_{01}-3C+C_S$\\
 & & (1527) & $-A(9C+C_S)$\\
$100[3]_{FS}[3]_F[3]_S$ & ${3\over 2}^+,\Delta(1600)$ & 1541-1675 &
$\epsilon_{10}-3C-C_S$\\
 & & (1674) & \\
$011[21]_{FS}[3]_F[21]_S$ & ${1\over 2}^-,\Delta(1620);{3\over
2}^-,\Delta(1700)$ & 1575-1700 & $\epsilon_{01}+3C+C_S$\\
 & & (1687) & $-A(9C+3C_S)$\\
$011[21]_{FS}[21]_F[3]_S$ & ${1\over 2}^-,N(1650);{3\over
2}^-,N(1700)$ & 1648-1710 & $\epsilon_{01}+3C-C_S$\\
 & ${5\over 2}^-,N(1675)$ & (1666) & $-A(9C-C_S)$\\
$022[3]_{FS}[3]_F[3]_S$ & ${1\over 2}^+;\Delta(1750),{3\over
2}^+,\Delta(1754?)$ & 1710-1780 & $\epsilon_{02}-3C-C_S$\\
 &${5\over 2}^+,\Delta(1754),{7\over
2}^+,\Delta(?)$ & (1832) & $+3A(3C+C_S)$\\ 
$022[3]_{FS}[21]_F[21]_S$ & ${3\over 2}^+,N(1720); {5\over
2}^+,N(1680)$ & 1656-1748 & $\epsilon_{02}-15C+C_S$\\
 & & (1731) & $+3A(15C-C_S)$ \\
$020[21]_{FS}[21]_F[21]_S$ & ${1\over 2}^+,N(1710)$ & 1636-1770 &
$\epsilon_{02}-3C+C_S$\\
 & & (1742) & $-A(9C+C_S)$ \\
$020[21]_{FS}[21]_F[3]_S$ & ${3\over 2}^+,N(1879?)$ & ? &
$\epsilon_{02}+3C-C_S$\\
 & & (1881) &$-A(9C-C_S)$ \\
$022[21]_{FS}[21]_F[21]_S$ & ${3\over 2}^+,N(1900);
{5\over 2}^+, N(2000)$ & 1879-2175 &
$\epsilon_{02}-3C+C_S$\\
 & & (1875) &$+{A\over 2}(27C-5C_S)$ \\
$022[21]_{FS}[21]_F[3]_S$ & ${1\over 2}^+,N(?);
{3\over 2}^+, N(?)$ & 1920-2114 &
$\epsilon_{02}+3C-C_S$\\
 &${5\over 2}^+N(?);{7\over 2}^+,N(1990)$ & 
(1915) &$-{A\over 2}(9C-5C_S)$ \\
$020[21]_{FS}[3]_F[21]_S$ & ${1\over 2}^+,\Delta(1910)$ & 1792-1950 &
$\epsilon_{02}+3C+C_S$\\
 & & (1902) &$-A(9C+3C_S)$ \\
$022[21]_{FS}[3]_F[21]_S$ & ${3\over 2}^+,\Delta(1920);
{5\over 2}^+,\Delta(1905)$ & 1794-1870 &
$\epsilon_{02}+3C+C_S$\\
 & & (1936) & $-{A\over 2}(9C+3C_S)$\\ \hline
\end{tabular}
\end{center}

\newpage
\centerline{\bf Table 4}

The $S$, $P$ and $SD$ shell states in the $\Lambda$ hyperon
spectrum.
The column
$\epsilon$ contains the eigenvalues of the mass operator.
The
averages over the multiplets of of the empirically
extracted mass values are denoted EXP. The model values
obtained with the mass operators (2.7) with the confining
well (2.18) and the hyperfine interaction (2.20) are
listed below the empirical ones. 

\begin{center}
\begin{tabular}{l|l|l|l} \hline
$nKL[f]_{FS}[f]_F[f]_S$ & $LS$ Multiplet & EXP & $\epsilon$
\\ 
 & & (model value) &$\sqrt{\epsilon_{nK}^2+\Delta_S^2}-c$\\ \hline
$000[3]_{FS}[21]_F[21]_S$ & ${1\over 2}^+, \Lambda$ & 1116 &
$\hat\epsilon_{00}-9C-5C_S$\\
 & & (1116) & \\
$011[21]_{FS}[111]_F[21]_S$ & ${1\over 2}^-, \Lambda(1405)$ & 
1405-1520 &
$\hat\epsilon_{01}-3C-5C_S$\\
 &${3\over 2}^-,\Lambda(1520)$ & (1535) & $-A(3C+5C_S)$\\
$100[3]_{FS}[21]_F[21]_S$ & ${1\over 2}^+,\Lambda(1600)$ & 1560-1700 &
$\hat\epsilon_{10}-9C-5C_S$\\
 & & (1524) & \\
$011[21]_{FS}[21]_F[21]_S$ & ${1\over 2}^-,\Lambda(1670)$ 
& 1660-1695 & $\hat\epsilon_{01}-3C+C_S$\\
 &${3\over2}^-\Lambda(1690)$ & (1640) & $-A(3C+7C_S)$\\
$011[21]_{FS}[21]_F[3]_S$ & ${1\over 2}^-,\Lambda(1800);{3\over
2}^-,\Lambda(?)$ & 1720-1830 & $\hat\epsilon_{01}+3C-C_S$\\
 & ${5\over 2}^-,\Lambda(1830)$ & (1773) & $-2A(3C+C_S)$\\
$020[21]_{FS}[111]_F[21]_S$ & ${1\over 2}^+,\Lambda(1810)$
 & 1750-1850 & $\hat\epsilon_{02}-3C-5C_S$\\
 & & (1739) & $-A(3C+5C_S)$\\
$022[3]_{FS}[21]_F[21]_S$ & ${3\over 2}^+,\Lambda(1890)$
 & 1815-1910 & $\hat\epsilon_{02}-9C-5C_S$\\
 &${5\over 2}^+,\Lambda(1820)$ & (1839) & $+3A(9C+5C_S)$\\
$020[21]_{FS}[21]_F[21]_S$ & ${1\over 2}^+,\Lambda(?)$ & ? &
$\hat\epsilon_{02}-3C+C_S$\\
 & & (1844) &$-A(3C+7C_S)$ \\
$020[21]_{FS}[21]_F[3]_S$ & ${3\over 2}^+,\Lambda(?)$ & ? &
$\hat\epsilon_{02}+3C-C_S$\\
 & & (1977) &$-2A(3C+C_S)$ \\
$022[21]_{FS}[21]_F[3]_S$ & ${1\over 2}^+, \Lambda(?);
{3\over 2}^+, \Lambda(?)$ & 2020? &
$\hat\epsilon_{02}+3C-C_S$\\
 &${5\over 2}^+  \Lambda(?);{7\over 2}^+,\Lambda(2020)$ & 
(2001) &$-2A(3C-2C_S)$ \\
$022[21]_{FS}[111]_F[21]_S$ & ${3\over 2}^+, \Lambda(?);
{5\over 2}^+, \Lambda(2110?)$ & 2090-2140 &
$\hat\epsilon_{02}-3C-5C_S$\\
 & & (1916) &$+{A\over 2}(21C+35C_S)$ \\
$022[21]_{FS}[21]_F[21]_S$ & ${3\over 2}^+, \Lambda(?);
{5\over 2}^+, \Lambda(2110?)$ & 2090-2140 & $\hat\epsilon_{02}-3C+C_S$\\
 & & (1959) &$+{A\over 2}(21C+C_S)$ \\ \hline
\end{tabular}
\end{center}

\newpage
\centerline{\bf Table 5}

The $S$, $P$ and $SD$ shell states in the $\Sigma$ hyperon
spectrum.
The column
$\epsilon$ contains the eigenvalues of the mass operator. The
averages over the multiplet of the empirically
extracted mass values are denoted EXP. The model values
obtained with the mass operators (2.7) with the confining
well (2.18) and the hyperfine interaction (2.20) are
listed below the empirical ones. The unconfirmed unassigned
low lying $\Sigma(1450)$ and $\Sigma(1560)$ resonances 
still included in ref. \cite{PDG} have been omitted from
this table.

\begin{center}
\begin{tabular}{l|l|l|l} \hline
$nKL[f]_{FS}[f]_F[f]_S$ & $LS$ Multiplet & EXP & $\epsilon$
\\ 
 & & (model value) &$\sqrt{\epsilon_{nK}^2+\Delta_S^2}-c$ \\ \hline
$000[3]_{FS}[21]_F[21]_S$ & ${1\over 2}^+, \Sigma $ & 
1189-1192 &
$\hat\epsilon_{00}-C-13 C_S$\\
 & & (1188) & \\
$000[3]_{FS}[3]_F[3]_S$ & ${3\over 2}^+,\Sigma(1385)$ & 1379 &
$\hat\epsilon_{00}-C-3C_S$\\
 & & (1378) & \\
$100[3]_{FS}[21]_F[21]_S$ & ${1\over 2}^+,\Sigma(1660)$ & 1660-1690 &
$\hat\epsilon_{10}-C-13 C_S$\\
 & & (1596) & \\
$011[21]_{FS}[21]_F[21]_S$ & ${1\over 2}^-,\Sigma(1620);{3\over
2}^-\Sigma(1580)$ & 1580-1640 & $\hat\epsilon_{01}+C-3 C_S$\\
 & & (1676) & $-A(3C+7 C_S)$\\
$100[3]_{FS}[3]_F[3]_S$ & ${3\over 2}^+,\Sigma(1840?)$ & 1720-1925 &
$\hat\epsilon_{10}-C-3C_S$\\
 & & (1786) & \\
$011[21]_{FS}[3]_F[21]_S$ & ${1\over 2}^-,\Sigma(?);{3\over
2}^-,\Sigma(1670)$ & 1665-1685 & $\hat\epsilon_{01}+C+3C_S$\\
 & & (1782) & $-A(3C+9C_S)$\\
$011[21]_{FS}[21]_F[3]_S$ & ${1\over 2}^-,\Sigma(1750);{3\over
2}^-,\Sigma(?)$ & 1730-1780 & $\hat\epsilon_{01}-C+3C_S$\\
 & ${5\over 2}^-,\Sigma(1775)$ & (1749) & $-8A C_S$\\
$022[3]_{FS}[3]_F[3]_S$ & ${1\over 2}^+,\Sigma(1770);{3\over
2}^+,\Sigma(?)$ & 1740-1790 & $\hat\epsilon_{02}-C-3C_S$\\
 &${5\over 2}^+,\Sigma(?);{7\over
2}^+,\Sigma(?)$ & (1928) & $+3A(C+3C_S)$\\ 
$022[3]_{FS}[21]_F[21]_S$ & ${3\over 2}^+,\Sigma(1840?);{5\over
2}^+,\Sigma(1915)$ & 1720-1937 & $\hat\epsilon_{02}-C-13C_S$\\
 & & (1864) & $+3A(C+13C_S)$\\
$020[21]_{FS}[21]_F[21]_S$ & ${1\over 2}^+,\Sigma(1880)$ & 1826-1985 &
$\hat\epsilon_{02}+C-3C_S$\\
 & & (1880) &$-A(3C+7C_S)$ \\
$020[21]_{FS}[21]_F[3]_S$ & ${3\over 2}^+,\Sigma(?)$ & ? &
$\hat\epsilon_{02}-C+3C_S$\\
 & & (1953) &$-8A C_S$ \\
$022[21]_{FS}[21]_F[21]_S$ & ${3\over 2}^+,\Sigma(?);
{5\over 2}^+, \Sigma(?)$ & ? &
$\hat\epsilon_{02}+C-3C_S$\\
 & & (1971) &$-{A\over 2}(3C-25C_S)$ \\
$022[21]_{FS}[21]_F[3]_S$ & ${1\over 2}^+,\Sigma(?);
{3\over 2}^+, \Sigma(?)$ & 2025-2040 &
$\hat\epsilon_{02}-C+3C_S$\\
 &${5\over 2}^+,\Sigma(?);{7\over 2}^+,\Sigma(2030)$ & 
(1984) &$+A (3C-5C_S)$ \\
$020[21]_{FS}[3]_F[21]_S$ & ${1\over 2}^+,\Sigma(?)$ & ? &
$\hat\epsilon_{02}+C+3C_S$\\
 & & (1986) &$-A(3C+9C_S)$ \\
$022[21]_{FS}[3]_F[21]_S$ & ${3\over 2}^+,\Sigma(2080);
{5\over 2}^+,\Sigma(2070)$ & 2051-2091 &
$\hat\epsilon_{02}+C+3C_S$\\
 & & (2014) & $-{A\over 2}(3C+9C_S)$\\ \hline
\end{tabular}
\end{center}

\newpage
\centerline{\bf Table 6}

The $S$, $P$ and $SD$ shell states in the $\Xi$ hyperon
spectrum.
The column
$\epsilon$ contains the eigenvalues of the mass operator. The
averages over the multiplets of the of the empirically
extracted mass values are denoted EXP. The model values
obtained with the mass operators (2.7) with the confining
well (2.18) and the hyperfine interaction (2.20) are
listed below the empirical ones. 

\begin{center}
\begin{tabular}{l|l|l|l} \hline
$nKL[f]_{FS}[f]_F[f]_S$ & $LS$ Multiplet & EXP & $\epsilon$
\\ 
 & & (model value) &$\sqrt{\epsilon_{nK}^2+2\Delta_S^2}-c$ \\ \hline
$000[3]_{FS}[21]_F[21]_S$ & ${1\over 2}^+, \Xi $ & 
1314-1321 &
$\hat\epsilon_{00}-14 C_S$\\
 & & (1310) & \\
$000[3]_{FS}[3]_F[3]_S$ & ${3\over 2}^+,\Xi(1530)$ & 1532-1534 &
$\hat\epsilon_{00}-4C_S$\\
 & & (1500) & \\
$100[3]_{FS}[21]_F[21]_S$ & ${1\over 2}^+,\Xi(1690?)$ & 1690 &
$\hat\epsilon_{10}-14 C_S$\\
 & & (1695) & \\
$011[21]_{FS}[21]_F[21]_S$ & ${1\over 2}^-,\Xi(?);{3\over
2}^-,\Xi(1820)$ & ? & $\hat\epsilon_{01}-2 C_S$\\
 & & (1769) & $ - 10 A C_S$\\
$100[3]_{FS}[3]_F[3]_S$ & ${3\over 2}^+,\Xi(?)$ & ? &
$\hat\epsilon_{10}-4C_S$\\
 & & (1885) & \\
$011[21]_{FS}[3]_F[21]_S$ & ${1\over 2}^-,\Xi(?);{3\over
2}^-,\Xi(?)$ & ? & $\hat\epsilon_{01}+4C_S$\\
 & & (1875) & $-12A C_S$\\
$011[21]_{FS}[21]_F[3]_S$ & ${1\over 2}^-,\Xi(?);{3\over
2}^-,\Xi(?)$ & ? & $\hat\epsilon_{01}+2C_S$\\
 & ${5\over 2}^-,\Xi(?)$ & (1853) & $-8A C_S$\\
$022[3]_{FS}[3]_F[3]_S$ & ${1\over 2}^+,\Xi(?);{3\over
2}^+,\Xi(?)$ & ? & $\hat\epsilon_{02}-4C_S$\\
 &${5\over 2}^+,\Xi(?),{7\over
2}^+,\Xi(?)$ & (2017) & $+12A C_S $\\ 
$022[3]_{FS}[21]_F[21]_S$ & ${3\over 2}^+,\Xi(?);{5\over
2}^+,\Xi(?)$ & ? & $\hat\epsilon_{02}-14C_S$\\
 & & (1953) & $ + 42A C_S $ \\
$020[21]_{FS}[21]_F[21]_S$ & ${1\over 2}^+,\Xi(?)$ & ? &
$\hat\epsilon_{02}-2C_S$\\
 & & (1963) & $ - 10 A C_S $ \\
$020[21]_{FS}[21]_F[3]_S$ & ${3\over 2}^+,\Xi$ & ? &
$\hat\epsilon_{02}+2C_S$\\
 & & (2048) &$-8A C_S$ \\
$022[21]_{FS}[21]_F[21]_S$ & ${3\over 2}^+,\Xi(?);
{5\over 2}^+, \Xi(?)$ & ? &
$\hat\epsilon_{02}-2C_S$\\
 & & (2051) &$+11A C_S$ \\
$022[21]_{FS}[21]_F[3]_S$ & ${1\over 2}^+,\Xi(?);
{3\over 2}^+, \Xi(?)$ & ? &
$\hat\epsilon_{02} +2C_S$\\
 &${5\over 2}^+,\Xi(?);{7\over 2}^+,\Xi$ & 
(2073) &$-2A C_S$ \\
$020[21]_{FS}[3]_F[21]_S$ & ${1\over 2}^+,\Xi(?)$ & ? &
$\hat\epsilon_{02}+4C_S$\\
 & & (2069) & $ - 12 A C_ S $ \\
$022[21]_{FS}[3]_F[21]_S$ & ${3\over 2}^+,\Xi(?);
{5\over 2}^+,\Xi(?)$ & ?&
$\hat\epsilon_{02}+4C_S$\\
 & & (2094) & $-6A C_S$\\ \hline
\end{tabular}
\end{center}

\newpage
\centerline{\bf Table 7 }

The $S$, $P$ and $SD$ shell states in the $\Omega^-$ hyperon
spectrum.
The column
$\hat\epsilon$ contains the eigenvalues  of 
the mass operator. The
averages over the multiplets of the empirically
extracted mass values are denoted EXP. The model values
obtained with the mass operators (2.7) with the confining
well (2.18) and the hyperfine interaction (2.20) are
listed below the empirical ones. 

\begin{center}
\begin{tabular}{l|l|l|l} \hline
$nKL[f]_{FS}[f]_F[f]_S$ & $LS$ Multiplet & EXP & $\epsilon$
\\ 
 & & (model value) &$\sqrt{\epsilon_{nK}^2+3\Delta_S^2}-c$ \\ \hline
$000[3]_{FS}[3]_F[3]_S$ & ${3\over 2}^+,\Omega^-$ & 1672  &
$\hat\epsilon_{00}-4C_S$\\
 & & (1606) & \\
$100[3]_{FS}[3]_F[3]_S$ & ${3\over 2}^+,\Omega^-(?)$ & ? &
$\hat\epsilon_{10}-4C_S$\\
 & & (1971) & \\
$011[21]_{FS}[3]_F[21]_S$ & ${1\over 2}^-,\Omega^-(?);{3\over
2}^-,\Omega^-(?)$ & ? & $\hat\epsilon_{01}+4C_S$\\
 & & (1953) & $ - 12 A C_S $ \\
$022[3]_{FS}[3]_F[3]_S$ & ${1\over 2}^+;\Omega^-(?),{3\over
2}^+,\Omega^-(?)$ & ? & $\hat\epsilon_{02}-4C_S$\\
 &${5\over 2}^+,\Omega^-(?),{7\over
2}^+,\Omega^-(?)$ & (2100) & $+12A C_S$\\ 
$020[21]_{FS}[3]_F[21]_S$ & ${1\over 2}^+\Omega^-(?)$ & ? &
$\hat\epsilon_{02}+4C_S$\\
 & & (2262) & $ - 12 A C_S  $ \\
$022[21]_{FS}[3]_F[21]_S$ & ${3\over 2}^+,\Omega^-(?);
{5\over 2}^+,\Omega^-(?)$ & ? &
$\hat\epsilon_{02}+4C_S$\\
 & & (2177) & $-6A C_S$\\ \hline
\end{tabular}
\end{center}

\newpage
\centerline{\bf Table 8}

The $S$, $P$ and $SD$ shell states in the $\Lambda_C^+$ hyperon
spectrum.
The column
$\epsilon$ contains the eigenvalues of the mass operator.
The
averages over the multiplets of of the empirically
extracted mass values are denoted EXP. The model values
obtained with the mass operators (2.7) with the confining
well (2.18) and the hyperfine interaction (2.20) are
listed below the empirical ones. 

\begin{center}
\begin{tabular}{l|l|l|l} \hline
$nKL[f]_{FS}[f]_F[f]_S$ & $LS$ Multiplet & EXP & $\epsilon$
\\ 
 & & (model value) &$\sqrt{\epsilon_{nK}^2+\Delta_C^2}-c$\\ \hline
$000[3]_{FS}[21]_F[21]_S$ & ${1\over 2}^+, \Lambda_C^+$ & 2285 &
$\hat\epsilon_{00}-9C+C_S-6C_C$\\
 & & (2285) & \\
$011[21]_{FS}[111]_F[21]_S$ & ${1\over 2}^-, \Lambda_C(2593)^+$ & 
2593-2625 &
$\hat\epsilon_{01}-3C+{1\over 3}C_S-4C_C$\\
 &${3\over 2}^-,\Lambda_C(2625)^+$ & (2610) & $-A(3C-
{1\over 3} C_S
+4C_C)$\\
$100[3]_{FS}[21]_F[21]_S$ & ${1\over 2}^+,\Lambda_C^+(?)$ & ? &
$\hat\epsilon_{10}-9C+C_S-6C_C$\\
 & & (2537) & \\
$011[21]_{FS}[21]_F[21]_S$ & ${1\over 2}^-,\Lambda_C^+(?))$ 
& ? & $\hat\epsilon_{01}-3C+{1\over 3}C_S+2C_C$\\
 &${3\over2}^-\Lambda_C^+(?)$ & (2656) & $-A(3C-{1\over 3}C_S
+10C_C)$\\
$011[21]_{FS}[21]_F[3]_S$ & ${1\over 2}^-,\Lambda_C^+(?);{3\over
2}^-,\Lambda_C^+(?)$ & ? & $\hat\epsilon_{01}+3C-{1\over 3}C_S
-2C_C$\\
 & ${5\over 2}^-,\Lambda_C^+(?)$ & (2772) & $-A(6C-
{2\over 3}C_S+2C_C)$\\
$020[21]_{FS}[111]_F[21]_S$ & ${1\over 2}^+,\Lambda_C^+(?)$
 & ? & $\hat\epsilon_{02}-3C+{1\over 3}C_S-4C_C$\\
 & & (2743) & $-A(3C-{1\over 3}C_S+4C_C)$\\
$022[3]_{FS}[21]_F[21]_S$ & ${3\over 2}^+,\Lambda_C^+(?)$
 & ? & $\hat\epsilon_{02}-9C+C_S-6C_S$\\
 &${5\over 2}^+,\Lambda_C^+(?)$ & (2787) & $+A(27C-3C_S+18C_C)$\\
$020[21]_{FS}[21]_F[21]_S$ & ${1\over 2}^+,\Lambda_C^+(?)$ & ? &
$\hat\epsilon_{02}-3C+{1\over 3}C_S+2C_C$\\
 & & (2790) &$-A(3C-{1\over 3}C_S+10C_C)$ \\
$020[21]_{FS}[21]_F[3]_S$ & ${3\over 2}^+,\Lambda_C^+(?)$ & ? &
$\hat\epsilon_{02}+3C-{1\over 3}C_S-2C_C$\\
 & & (2906) &$-A(6C-{2\over 3}C_S+2C_C)$ \\
$022[21]_{FS}[21]_F[3]_S$ & ${1\over 2}^+, \Lambda_C^+(?);
{3\over 2}^+, \Lambda_C^+(?)$ & ? &
$\hat\epsilon_{02}+3C-{1\over 3}C_S-2C_C$\\
 &${5\over 2}^+  \Lambda_C^+(?);{7\over 2}^+,\Lambda_C^+(?)$ & 
(2926) &$-A(6C-{2\over 3}C_S-7C_C)$ \\
$022[21]_{FS}[111]_F[21]_S$ & ${3\over 2}^+, \Lambda_C^+(?);
{5\over 2}^+, \Lambda_C^+(?)$ & ? &
$\hat\epsilon_{02}-3C+{1\over 3}C_S-4C_C$\\
 & & (2858) &$+A({21\over 2}C-{7\over 6}C_S+14C_C)$ \\
$022[21]_{FS}[21]_F[21]_S$ & ${3\over 2}^+, \Lambda_C^+(?);
{5\over 2}^+, \Lambda_C^+(?)$ & ? & $\hat\epsilon_{02}-3C+
{1\over 3}C_S+2C_C$\\
 & & (2885) &$+A({21\over 2}C-{7\over 6}C_S-C_C)$ \\ \hline
\end{tabular}
\end{center}

\newpage
\centerline{\bf Table 9}

The $S$, $P$ and $SD$ shell states in the $\Sigma_C$ hyperon
spectrum.
The column
$\epsilon$ contains the eigenvalues of the mass operator. The
averages over the multiplet of the empirically
extracted mass values are denoted EXP. The model values
obtained with the mass operators (2.7) with the confining
well (2.18) and the hyperfine interaction (2.20) are
listed below the empirical ones.

\begin{center}
\begin{tabular}{l|l|l|l} \hline
$nKL[f]_{FS}[f]_F[f]_S$ & $LS$ Multiplet & EXP & $\epsilon$
\\ 
 & & (model value) &$\sqrt{\epsilon_{nK}^2+\Delta_C^2}-c$ \\ \hline
$000[3]_{FS}[21]_F[21]_S$ & ${1\over 2}^+, \Sigma_C $ & 
2452-2455 &
$\hat\epsilon_{00}-C-{1\over 3}C_S-10C_C$\\
 & & (2444) & \\
$000[3]_{FS}[3]_F[3]_S$ & ${3\over 2}^+,\Sigma_C(2530)$ & 2518 &
$\hat\epsilon_{00}-C-{1\over 3}C_S-4C_C$\\
 & & (2504) & \\
$100[3]_{FS}[21]_F[21]_S$ & ${1\over 2}^+,\Sigma_C(?)$ & ? &
$\hat\epsilon_{10}-C-{1\over 3}C_S-10 C_C$\\
 & & (2696) & \\
$011[21]_{FS}[21]_F[21]_S$ & ${1\over 2}^-,\Sigma_C(?);{3\over
2}^-\Sigma_C(?)$ & ? & $\hat\epsilon_{01}+C+{1\over 3} C_S
-2C_C$\\
 & & (2732) & $-A(3C+C_S+6C_C)$\\
$100[3]_{FS}[3]_F[3]_S$ & ${3\over 2}^+,\Sigma_C(?)$ & ? &
$\hat\epsilon_{10}-C-{1\over 3}C_S-4C_C$\\
 & & (2754) & \\
$011[21]_{FS}[3]_F[21]_S$ & ${1\over 2}^-,\Sigma_C(?);{3\over
2}^-,\Sigma_C(?)$ & ? & $\hat\epsilon_{01}+C+{1\over 3}C_S
+4C_C$\\
 & & (2778) & $-A(3C+C_S+12C_C)$\\
$011[21]_{FS}[21]_F[3]_S$ & ${1\over 2}^-,\Sigma_C(?);{3\over
2}^-,\Sigma_C(?)$ & ? & $\hat\epsilon_{01}-C-{1\over 3}C_S
+2C_C$\\
 & ${5\over 2}^-,\Sigma_C(?)$ & (2724) & $-6A C_C$\\
$022[3]_{FS}[3]_F[3]_S$ & ${1\over 2}^+,\Sigma_C(?);{3\over
2}^+,\Sigma_C(?)$ & ? & $\hat\epsilon_{02}-C-{1\over 3}C_S
-4C_C$\\
 &${5\over 2}^+,\Sigma_C(?);{7\over
2}^+,\Sigma_C(?)$ & (2862) & $+A(3C+C_S+12C_C)$\\ 
$022[3]_{FS}[21]_F[21]_S$ & ${3\over 2}^+,\Sigma_C(?);{5\over
2}^+,\Sigma_C(?)$ & ? & $\hat\epsilon_{02}-C-{1\over 3}C_S
-10C_C$\\
 & & (2841) & $+A(3C+C_S+30C_C)$\\
$020[21]_{FS}[21]_F[21]_S$ & ${1\over 2}^+,\Sigma_C(?)$ & ? &
$\hat\epsilon_{02}+C+{1\over 3}C_S-2C_C$\\
 & & (2865) &$-A(3C+C_S+6C_C)$ \\
$020[21]_{FS}[21]_F[3]_S$ & ${3\over 2}^+,\Sigma_C(?)$ & ? &
$\hat\epsilon_{02}-C-{1\over 3}C_S+2C_C$\\
 & & (2859) &$ - 6 A C_C $ \\
$022[21]_{FS}[21]_F[21]_S$ & ${3\over 2}^+,\Sigma_C(?);
{5\over 2}^+, \Sigma_C(?)$ & ? &
$\hat\epsilon_{02}+C+{1\over 3}C_S-2C_C$\\
 & & (2908) &$-A({3\over 2}C+{1\over 2}C_S
-9C_C)$ \\
$022[21]_{FS}[21]_F[3]_S$ & ${1\over 2}^+,\Sigma_C(?);
{3\over 2}^+, \Sigma_C(?)$ & ? &
$\hat\epsilon_{02}-C-{1\over 3}C_S+2C_C$\\
 &${5\over 2}^+,\Sigma_C(?);{7\over 2}^+,\Sigma_C(?)$ & 
(2888) &$+A (3C+C_S-3C_C)$ \\
$020[21]_{FS}[3]_F[21]_S$ & ${1\over 2}^+,\Sigma_C(?)$ & ? &
$\hat\epsilon_{02}+C+{1\over 3}C_S+4C_C$\\
 & & (2912) &$-A(3C+C_S+12C_C)$ \\
$022[21]_{FS}[3]_F[21]_S$ & ${3\over 2}^+,\Sigma_C(?);
{5\over 2}^+,\Sigma_C(?)$ & ? &
$\hat\epsilon_{02}+C+{1\over 3}C_S+4C_C$\\
 & & (2936) & $-A({3\over 2}C+{1\over 2}C_S+6C_C)$\\ \hline
\end{tabular}
\end{center}

\newpage
\centerline{\bf Table 10}

The $S$, $P$ and $SD$ shell states in the $\Xi_C^a$ hyperon
spectrum.
The column
$\epsilon$ contains the eigenvalues  of the mass operator.
The
averages over the multiplets of of the empirically
extracted mass values are denoted EXP. The model values
obtained with the mass operators (2.7) with the confining
well (2.18) and the hyperfine interaction (2.20) are
listed below the empirical ones. 

\begin{center}
\begin{tabular}{l|l|l|l} \hline
$nKL[f]_{FS}[f]_F[f]_S$ & $LS$ Multiplet & EXP & $\epsilon$
\\ 
 & & (model value) 
&$\sqrt{\epsilon_{nK}^2+\Delta_S^2+\Delta_C^2}$ \\ \hline
$000[3]_{FS}[21]_F[21]_S$ & ${1\over 2}^+, \Xi_C^a$ & 2465-2470 &
$\hat\epsilon_{00}-8C_S-3C_C-3C_{SC}$\\
 & & (2447) & \\
$011[21]_{FS}[111]_F[21]_S$ & ${1\over 2}^-, \Xi_C^a(?)$ & 
? &
$\hat\epsilon_{01}-{8\over3}C_{S}-2C_C-2C_{SC}$\\
 &${3\over 2}^-,\Xi_C^a$ 
& (2716) & $-A({8\over3} C_{S}+2C_C+2C_{SC})$\\
$100[3]_{FS}[21]_F[21]_S$ & ${1\over 2}^+,\Xi_C^a(?)$ & ? &
$\hat\epsilon_{10}-8C_S-3C_{C}-3C_{SC}$\\
 & & (2693) & \\
$011[21]_{FS}[21]_F[21]_S$ & ${1\over 2}^-,\Xi_C^a(?)$ 
& ? & $\hat\epsilon_{01}-{8\over3}C_{S}+C_C+C_{SC}$\\
 &${3\over2}^-\Xi_C^a(?)$ 
& (2752) & $-A({8\over3}C_{S}+5C_C+5C_{SC})$\\
$011[21]_{FS}[21]_F[3]_S$ & ${1\over 2}^-,\Xi_C^a(?);{3\over
2}^-,\Xi_C^a(?)$ &2814- 2819 & $\hat\epsilon_{01}
+{8\over3}C_{S}-C_{C}-C_{SC}$\\
 & ${5\over 2}^-,\Xi_C^a(?)$ & (2825) & $-A({16\over3}C_{S}
+C_C+C_{SC})$\\
$020[21]_{FS}[111]_F[21]_S$ & ${1\over 2}^+,\Xi_C^a(?)$
 & ? & $\hat\epsilon_{02}-{8\over 3}C_S-2C_C-2C_{SC}$\\
 & & (2847) & $-2A({4\over 3}C_S+C_C+C_{SC})$\\
$022[3]_{FS}[21]_F[21]_S$ & ${3\over 2}^+,\Xi_C^a(?)$
 & ? & $\hat\epsilon_{02}-8C_S-3C_C-3C_{SC}$\\
 &${5\over 2}^+,\Xi_C^a(?)$ & (2879) & $+3A(8C_S+3C_S+3C_{SC})$\\
$020[21]_{FS}[21]_F[21]_S$ & ${1\over 2}^+,\Xi_C^a(?)$ & ? &
$\hat\epsilon_{02}-{8\over 3}C_S+C_C+C_{SC}$\\
 & & (2882) &$-A({8\over 3}C_S+5C_C+5C_{SC})$ \\
$020[21]_{FS}[21]_F[3]_S$ & ${3\over 2}^+,\Xi_C^a(?)$ & ? &
$\hat\epsilon_{02}+{8\over 3}C_S-C_C-C_{SC}$\\
 & & (2977) &$-A({16\over 3}C_S+C_C+C_{SC})$ \\
$022[21]_{FS}[21]_F[3]_S$ & ${1\over 2}^+,\Xi_C^a(?);
{3\over 2}^+, \Xi_C^a(?)$ & ? &
$\hat\epsilon_{02}+{8\over 3}C_S-C_C-C_{SC}$\\
 &${5\over 2}^+,  \Xi_C^a(?);{7\over 2}^+,\Xi_C^a(?)$ & 
(2970) &$-A({16\over 3}C_S-{7\over 2}C_C-{7\over 2}C_{SC})$ \\
$022[21]_{FS}[111]_F[21]_S$ & ${3\over 2}^+,\Xi_C^a(?);
{5\over 2}^+, \Xi_C^a(?)$ & ? &
$\hat\epsilon_{02}-{8\over 3}C_S-2C_C-2C_{SC}$\\
 & & (2927) &$+A({28\over 3}C_S+7C_C+7C_{SC})$ \\
$022[21]_{FS}[21]_F[21]_S$ & ${3\over 2}^+,\Xi_C^a(?);
{5\over 2}^+,\Xi_C^a(?)$ & ? & $\hat\epsilon_{02}+{4\over 3}C_S
-C_C-C_{SC}$\\
 & & (2987) &$-A (2C_S-12C_C-12C_{SC})$ \\ \hline
\end{tabular}
\end{center}

\newpage
\centerline{\bf Table 11}

The $S$, $P$ and $SD$ shell states in the $\Xi^s$ hyperon
spectrum.
The column
$\epsilon$ contains the eigenvalues of the mass operator. The
averages over the multiplet of the empirically
extracted mass values are denoted EXP. The model values
obtained with the mass operators (2.7) with the confining
well (2.18) and the hyperfine interaction (2.20) are
listed below the empirical ones.

\begin{center}
\begin{tabular}{l|l|l|l} \hline
$nKL[f]_{FS}[f]_F[f]_S$ & $LS$ Multiplet & EXP & $\epsilon$
\\ 
 & & (model value) & $\sqrt{\epsilon_{nK}^2+
\Delta_ S^2+ \Delta^2_C }-c$\\ \hline
$000[3]_{FS}[21]_F[21]_S$ & ${1\over 2}^+, \Xi_C^s $ & 
2560 &
$\hat\epsilon_{00}-{4\over3} C_{S}-5C_C-5C_{SC}$\\
 & & (2543) & \\
$000[3]_{FS}[3]_F[3]_S$ & ${3\over 2}^+,\Xi_C^s$ & 2642 &
$\hat\epsilon_{00}-{4\over3}C_{S}-2C_C-2C_{SC}$\\
 & & (2589) & \\
$100[3]_{FS}[21]_F[21]_S$ & ${1\over 2}^+,\Xi_C^s(?)$ & ? &
$\hat\epsilon_{10}-{4\over3}C_{S}-5 C_S-5C_{SC}$\\
 & & (2789) & \\
$011[21]_{FS}[21]_F[21]_S$ & ${1\over 2}^-,\Xi_C^s(?);{3\over
2}^-\Xi_C^s(?)$ & ? & $\hat\epsilon_{01}+{4\over3}C_{S}- C_{C}-C_{SC}$\\
 & & (2798) & $-A(4C_{S}+3C_C+3C_{SC})$\\
$100[3]_{FS}[3]_F[3]_S$ & ${3\over 2}^+,\Xi_C^s(?)$ & ? &
$\hat\epsilon_{10}-{4\over3}C_{S}-2C_S-2C_{SC}$\\
 & & (2835) & \\
$011[21]_{FS}[3]_F[21]_S$ & ${1\over 2}^-,\Xi_C^s(?);{3\over
2}^-,\Xi_C^s(?)$ & ? & $\hat\epsilon_{01}+{4\over3}C_S+2C_C+2C_{SC}$\\
 & & (2833) & $-A(4C_S+C_C+6C_{SC})$\\
$011[21]_{FS}[21]_F[3]_S$ & ${1\over 2}^-,\Xi_C^s(?);{3\over
2}^-,\Xi_C^s(?)$ & ? & $\hat\epsilon_{01}-{4\over3}C_S+C_C+C_{SC}$\\
 & ${5\over 2}^-,\Xi_C^s(?)$ & (2795) & $-A(3C_{C}+3C_{SC}$\\
$022[3]_{FS}[3]_F[3]_S$ & ${1\over 2}^+,\Xi_C^s(?);{3\over
2}^+,\Xi_C^s(?)$ & ? & $\hat\epsilon_{02}-{4\over 3}C_S-2C_C
-2C_{SC}$\\
 &${5\over 2}^+,\Xi_C^s(?);{7\over
2}^+,\Xi_C^s(?)$ & (2927) & $+3A({4\over 3}C_S+2C_S+2C_{SC})$\\ 
$022[3]_{FS}[21]_F[21]_S$ & ${3\over 2}^+,\Xi_C^s(?);{5\over
2}^+,\Xi_C^s(?)$ & ? & $\hat\epsilon_{02}-{4\over 3}C_S
-5C_C-5C_{SC}$\\
 & & (2912) & $+3A({4\over 3}C_S+5C_C+5C_{SC})$\\
$020[21]_{FS}[21]_F[21]_S$ & ${1\over 2}^+,\Xi_C^s(?)$ & ? &
$\hat\epsilon_{02}+{4\over 3}C_S+3C_C+3C_{SC}$\\
 & & (2990) &$-A(4C_S+3C_S+3C_{SC})$ \\
$020[21]_{FS}[21]_F[3]_S$ & ${3\over 2}^+,\Xi_C^s(?)$ & ? &
$\hat\epsilon_{02}-{4\over 3}C_S+C_C+C_{SC}$\\
 & & (2926) &$-A(3C_C+3C_{SC})$ \\
$022[21]_{FS}[21]_F[21]_S$ & ${3\over 2}^+,\Xi_C^s(?);
{5\over 2}^+, \Xi_C^s(?)$ & ? &
$\hat\epsilon_{02}+{4\over 3}C_S-C_S-C_{SC}$\\
 & & (2963) &$-A(2C_S-{9\over 2}C_C-{9\over 2}C_{SC})$ \\
$022[21]_{FS}[21]_F[3]_S$ & ${1\over 2}^+,\Xi_C^s(?);
{3\over 2}^+, \Xi_C^s(?)$ & ? &
$\hat\epsilon_{02}-{4\over 3}C_S-2C_C-2C_{SC}$\\
 &${5\over 2}^+,\Xi_C^s(?);{7\over 2}^+,\Xi_C^s(?)$ & 
(2902) &$+A (4C_S-{3\over 2}C_C-{3\over 2}C_{SC})$ \\
$020[21]_{FS}[3]_F[21]_S$ & ${1\over 2}^+,\Xi_C^s(?)$ & ? &
$\hat\epsilon_{02}+{4\over 3}C_S+2C_C+2C_{SC}$\\
 & & (2952) &$-2A(2C_S+2C_C+2C_{SC})$ \\
$022[21]_{FS}[3]_F[21]_S$ & ${3\over 2}^+,\Xi_C^s(?);
{5\over 2}^+,\Xi_C^s(?)$ & ? &
$\hat\epsilon_{02}+{4\over 3}C_S+2C_C+2C_{SC}$\\
 & & (2989) & $-A(2C_S+C_C+C_{SC})$\\ \hline
\end{tabular}
\end{center}

\newpage
\centerline{\bf Table 12}

The $S$, $P$ and $SD$ shell states in the $\Omega_C^0$ hyperon
spectrum.
The column
$\epsilon$ contains the eigenvalues of the mass operator. The
averages over the multiplets of the of the empirically
extracted mass values are denoted EXP. The model values
obtained with the mass operators (2.7) with the confining
well (2.18) and the hyperfine interaction (2.20) are
listed below the empirical ones. 

\begin{center}
\begin{tabular}{l|l|l|l} \hline
$nKL[f]_{FS}[f]_F[f]_S$ & $LS$ Multiplet & EXP & $\epsilon$
\\ 
 & & (model value) &$\sqrt{\epsilon_{nK}^2+2\Delta_S^2+\Delta_C^2}
-c$\\ \hline
$000[3]_{FS}[21]_F[21]_S$ & ${1\over 2}^+, \Omega_C^0 $ & 
2710 &
$\hat\epsilon_{00}-{4\over3}C_{S}-10 C_{SC}$\\
 & & (2633) & \\
$000[3]_{FS}[3]_F[3]_S$ & ${3 \over 2}^+,\Omega_C^0(?)$ & ? &
$\hat\epsilon_{00}-{4\over 3}C_S-4C_{SC}$\\
 & & (2663) & \\
$100[3]_{FS}[21]_F[21]_S$ & ${1\over2}^+,\Omega_C^0(?)$ & ? &
$\hat\epsilon_{10}-{4\over3}C_{S}-10_{SC}$\\
 & & (2874) & \\
$011[21]_{FS}[21]_F[21]_S$ & ${1\over2}^-,\Omega_C^0(?);{3\over
2}^-,\Omega_C^0$ & ? & $\hat\epsilon_{01}+{4\over3}C_{S}-2C_{SC}$\\
 & & (2867) & $ -2 A(2C_S+3C_{SC})$\\
$100[3]_{FS}[3]_F[3]_S$ & ${ 3 \over 2}^+,\Omega_C^0(?)$ & ? &
$\hat\epsilon_{10}-{4\over3}C_{S}-4C_{SC}$\\
 & & (2904) & \\
$011[21]_{FS}[3]_F[21]_S$ & ${ 1\over 2}^-,\Omega_C^0;{3\over
2}^-,\Omega_C^0(?)$ & ? & $\hat\epsilon_{01}+{4\over3}C_S+4C_{SC}$\\
 & & (2890) & $-2A(2C_{S}+6C_{SC})$\\
$011[21]_{FS}[21]_F[3]_S$ & ${1 \over 2}^-,\Omega_C^0;{3\over
2}^-,\Omega_C^0(?)$ & ? & $\hat\epsilon_{01}-{4\over3}C_{S}+2C_{SC}$\\
 & ${5\over 2}^-,\Omega_C^0(?)$ & (2853) & $-6 A C_{SC}$\\
$022[3]_{FS}[3]_F[3]_S$ & ${1 \over 2}^+,\Omega^{0}_{C}(?);{3\over
2}^+,\Omega_C^0$ & ? & $\hat\epsilon_{02}-{4\over3}C_S-4C_{SC}$\\
 &${5\over 2}^+,\Omega_C^0(?),{7\over
2}^+,\Omega^{0}_{C}(?)$ & (2987) & $+6A
({2\over3}C_{C}+2C_{SC})$\\ 
$022[3]_{FS}[21]_F[21]_S$ & ${3 \over 2}^+,\Omega_C^0(?);{5\over
2}^+,\Omega_C^0(?)$ & ? & $\hat \epsilon_{02}-{4\over3}C_S-10C_{SC}$\\
 & & (2977) & $ +A (4C_{S}+30C_{SC}) $ \\
$020[21]_{FS}[21]_F[21]_S$ & ${1 \over 2}^+,\Omega_C^0(?)$ & ? &
$\hat\epsilon_{02}+{4\over3}C_S-2C_{SC}$\\
 & & (2996) & $ -2 A(2C_S+3C_{SC}) $ \\
$020[21]_{FS}[21]_F[3]_S$ & ${3 \over 2}^+,\Omega_C^0$ & ? &
$\hat\epsilon-{4\over3}C_{S}+2C_{SC}$\\
 & & (2981) &$-6 A C_{SC}$ \\
$022[21]_{FS}[21]_F[21]_S$ & ${3\over 2}^+,\Omega_C^0(?);
{5\over 2}^+, \Omega_C^0(?)$ & ? &
$\hat\epsilon_{02}+{4\over 3} C_{S}-2C_{SC}$\\
 & & (3019) &$- A (2C_{S}-9C_{SC})$ \\
$022[21]_{FS}[21]_F[3]_S$ & ${1 \over 2}^+,\Omega_C^0(?);
{3\over 2}^+, \Omega_C^0(?)$ & ? &
$\hat\epsilon_{02} -{4\over3}C_{S}+2C_{S}$\\
 &${5\over 2}^+,\Omega_C^0(?);{7 \over 2}^+,\Omega_C^0(?)$ & 
(3002) &$+A(4C_{S}-3C_{SC})$ \\
$020[21]_{FS}[3]_F[21]_S$ & ${1 \over 2}^+,\Omega_C^0(?)$ & ? &
$\hat\epsilon_{02}+{4\over3}C_{S}+4C_{SC}$\\
 & & (3019) & $ -2 A (2C_ S+6C_{SC})$ \\
$022[21]_{FS}[3]_F[21]_S$ & ${3\over 2}^+,\Omega_C^0(?);
{5\over 2}^+,\Omega_C^0(?)$ & ?&
$\hat\epsilon_{02}+{4\over3}C_S+4C_{SC}$\\
 & & (3040) & $-A(2C_{S}+6C_{SC})$\\ \hline
\end{tabular}
\end{center}

\newpage
\centerline{\bf Table 13}

The $S$, $P$ and $SD$ shell states in the $\Lambda_b^0$ hyperon
spectrum.
The column
$\epsilon$ contains the eigenvalues of the mass operator.
The
averages over the multiplets of of the empirically
extracted mass values are denoted EXP. The model values
obtained with the mass operators (2.7) with the confining
well (2.18) and the hyperfine interaction (2.20) are
listed below the empirical ones. 

\begin{center}
\begin{tabular}{l|l|l|l} \hline
$nKL[f]_{FS}[f]_F[f]_S$ & $LS$ Multiplet & EXP & $\epsilon$
\\ 
 & & (model value) &$\sqrt{\epsilon_{nK}^2+\Delta_B^2}-c$\\ \hline
$000[3]_{FS}[21]_F[21]_S$ & ${1\over 2}^+, \Lambda_b^0$ & 5641 &
$\hat\epsilon_{00}-9C+C_S-6C_B$\\
 & & (5641) & \\
$011[21]_{FS}[111]_F[21]_S$ & ${1\over 2}^-, \Lambda_b^0(?)$ & 
? &
$\hat\epsilon_{01}-3C+{1\over 3}C_S-4C_B$\\
 &${3\over 2}^-,\Lambda_b^0(?)$ & (5867) & $-A(3C-
{1\over 3} C_S
+4C_B)$\\
$100[3]_{FS}[21]_F[21]_S$ & ${1\over 2}^+,\Lambda_b^0(?)$ & ? &
$\hat\epsilon_{10}-9C+C_S-6C_B$\\
 & & (5754) & \\
$011[21]_{FS}[21]_F[21]_S$ & ${1\over 2}^-,\Lambda_b^0(?))$ 
& ? & $\hat\epsilon_{01}-3C+{1\over 3}C_S+2C_B$\\
 &${3\over2}^-\Lambda_b^0(?)$ & (5909) & $-A(3C-{1\over 3}C_S
+10C_B)$\\
$011[21]_{FS}[21]_F[3]_S$ & ${1\over 2}^-,\Lambda_b^0(?);{3\over
2}^-,\Lambda_b^0(?)$ & ? & $\hat\epsilon_{01}+3C-{1\over 3}C_S
-2C_B$\\
 & ${5\over 2}^-,\Lambda_b^0(?)$ & (6027) & $-A(6C-
{2\over 3}C_S+2C_B)$\\
$020[21]_{FS}[111]_F[21]_S$ & ${1\over 2}^+,\Lambda_b^0(?)$
 & ? & $\hat\epsilon_{02}-3C+{1\over 3}C_S-4C_B$\\
 & & (5930) & $-A(3C-{1\over 3}C_S+4C_B)$\\
$022[3]_{FS}[21]_F[21]_S$ & ${3\over 2}^+,\Lambda_b^0(?)$
 & ? & $\hat\epsilon_{02}-9C+C_S-6C_B$\\
 &${5\over 2}^+,\Lambda_b^0(?)$ & (5971) & $+A(27C-3C_S+18C_B)$\\
$020[21]_{FS}[21]_F[21]_S$ & ${1\over 2}^+,\Lambda_b^0(?)$ & ? &
$\hat\epsilon_{02}-3C+{1\over 3}C_S+2C_B$\\
 & & (5972) &$-A(3C-{1\over 3}C_S+10C_B)$ \\
$020[21]_{FS}[21]_F[3]_S$ & ${3\over 2}^+,\Lambda_b^0(?)$ & ? &
$\hat\epsilon_{02}+3C-{1\over 3}C_S-2C_B$\\
 & & (6090) &$-A(6C-{2\over 3}C_S+2C_B)$ \\
$022[21]_{FS}[21]_F[3]_S$ & ${1\over 2}^+, \Lambda_b^0(?);
{3\over 2}^+, \Lambda_b^0(?)$ & ? &
$\hat\epsilon_{02}+3C-{1\over 3}C_S-2C_B$\\
 &${5\over 2}^+  \Lambda_b^0(?);{7\over 2}^+,\Lambda_b^0(?)$ & 
(6108) &$-A(6C-{2\over 3}C_S-7C_B)$ \\
$022[21]_{FS}[111]_F[21]_S$ & ${3\over 2}^+, \Lambda_b^0(?);
{5\over 2}^+, \Lambda_b^0(?)$ & ? &
$\hat\epsilon_{02}-3C+{1\over 3}C_S-4C_B$\\
 & & (6042) &$+A({21\over 2}C-{7\over 6}C_S+14C_B)$ \\
$022[21]_{FS}[21]_F[21]_S$ & ${3\over 2}^+, \Lambda_B^0(?);
{5\over 2}^+, \Lambda_b^0(?)$ & ? & $\hat\epsilon_{02}-3C+
{1\over 3}C_S+2C_B$\\
 & & (6066) &$+A({21\over 2}C-{7\over 6}C_S-C_B)$ \\ \hline
\end{tabular}
\end{center}

\newpage
\centerline{\bf Table 14}

The $S$, $P$ and $SD$ shell states in the $\Sigma_b$ hyperon
spectrum.
The column
$\epsilon$ contains the eigenvalues of the mass operator. The
averages over the multiplet of the empirically
extracted mass values are denoted EXP. The model values
obtained with the mass operators (2.7) with the confining
well (2.18) and the hyperfine interaction (2.20) are
listed below the empirical ones.

\begin{center}
\begin{tabular}{l|l|l|l} \hline
$nKL[f]_{FS}[f]_F[f]_S$ & $LS$ Multiplet & EXP & $\epsilon$
\\ 
 & & (model value) &$\sqrt{\epsilon_{nK}^2+\Delta_B^2}-c$ \\ \hline
$000[3]_{FS}[21]_F[21]_S$ & ${1\over 2}^+, \Sigma_b $ & 
5814 &
$\hat\epsilon_{00}-C-{1\over 3}C_S-10C_B$\\
 & & (5801) & \\
$000[3]_{FS}[3]_F[3]_S$ & ${3\over 2}^+,\Sigma_b(?)$ & 5870 &
$\hat\epsilon_{00}-C-{1\over 3}C_S-4C_B$\\
 & & (5857) & \\
$100[3]_{FS}[21]_F[21]_S$ & ${1\over 2}^+,\Sigma_b(?)$ & ? &
$\hat\epsilon_{10}-C-{1\over 3}C_S-10 C_B$\\
 & & (5916) & \\
$011[21]_{FS}[21]_F[21]_S$ & ${1\over 2}^-,\Sigma_b(?);{3\over
2}^-\Sigma_b(?)$ & ? & $\hat\epsilon_{01}+C+{1\over 3} C_S
-2C_B$\\
 & & (5987) & $-A(3C+C_S+6C_B)$\\
$100[3]_{FS}[3]_F[3]_S$ & ${3\over 2}^+,\Sigma_b(?)$ & ? &
$\hat\epsilon_{10}-C-{1\over 3}C_S-4C_B$\\
 & & (5971) & \\
$011[21]_{FS}[3]_F[21]_S$ & ${1\over 2}^-,\Sigma_b(?);{3\over
2}^-,\Sigma_b(?)$ & ? & $\hat\epsilon_{01}+C+{1\over 3}C_S
+4C_B$\\
 & & (6030) & $-A(3C+C_S+12C_B)$\\
$011[21]_{FS}[21]_F[3]_S$ & ${1\over 2}^-,\Sigma_b(?);{3\over
2}^-,\Sigma_b(?)$ & ? & $\hat\epsilon_{01}-C-{1\over 3}C_S
+2C_B$\\
 & ${5\over 2}^-,\Sigma_b(?)$ & (5977) & $-6A C_B$\\
$022[3]_{FS}[3]_F[3]_S$ & ${1\over 2}^+,\Sigma_b(?);{3\over
2}^+,\Sigma_b(?)$ & ? & $\hat\epsilon_{02}-C-{1\over 3}C_S
-4C_B$\\
 &${5\over 2}^+,\Sigma_b(?);{7\over
2}^+,\Sigma_b(?)$ & (6044) & $+A(3C+C_S+12C_B)$\\ 
$022[3]_{FS}[21]_F[21]_S$ & ${3\over 2}^+,\Sigma_b(?);{5\over
2}^+,\Sigma_b(?)$ & ? & $\hat\epsilon_{02}-C-{1\over 3}C_S
-10C_B$\\
 & & (6026) & $+A(3C+C_S+30C_B)$\\
$020[21]_{FS}[21]_F[21]_S$ & ${1\over 2}^+,\Sigma_b(?)$ & ? &
$\hat\epsilon_{02}+C+{1\over 3}C_S-2C_B$\\
 & & (6050) &$-A(3C+C_S+6C_B)$ \\
$020[21]_{FS}[21]_F[3]_S$ & ${3\over 2}^+,\Sigma_b(?)$ & ? &
$\hat\epsilon_{02}-C-{1\over 3}C_S+2C_B$\\
 & & (6040) &$ - 6 A C_B $ \\
$022[21]_{FS}[21]_F[21]_S$ & ${3\over 2}^+,\Sigma_b(?);
{5\over 2}^+, \Sigma_b(?)$ & ? &
$\hat\epsilon_{02}+C+{1\over 3}C_S-2C_B$\\
 & & (6091) &$-A({3\over 2}C+{1\over 2}C_S
-9C_B)$ \\
$022[21]_{FS}[21]_F[3]_S$ & ${1\over 2}^+,\Sigma_b(?);
{3\over 2}^+, \Sigma_b(?)$ & ? &
$\hat\epsilon_{02}-C-{1\over 3}C_S+2C_B$\\
 &${5\over 2}^+,\Sigma_b(?);{7\over 2}^+,\Sigma_b(?)$ & 
(6069) &$+A (3C+C_S-3C_B)$ \\
$020[21]_{FS}[3]_F[21]_S$ & ${1\over 2}^+,\Sigma_b(?)$ & ? &
$\hat\epsilon_{02}+C+{1\over 3}C_S+4C_B$\\
 & & (6092) &$-A(3C+C_S+12C_B)$ \\
$022[21]_{FS}[3]_F[21]_S$ & ${3\over 2}^+,\Sigma_b(?);
{5\over 2}^+,\Sigma_b(?)$ & ? &
$\hat\epsilon_{02}+C+{1\over 3}C_S+4C_B$\\
 & & (6115) & $-A({3\over 2}C+{1\over 2}C_S+6C_B)$\\ \hline
\end{tabular}
\end{center}
\vspace{1cm}
\newpage
{\bf Figure Captions}
\vspace{0.5cm}

Fig. 1. The ground state wave functions $u_{00}(R)$ for the
confining models (2.17) (oscillator) and (2.18) (funnel) 
as functions of the
radius $R$. The wave functions have been calculated with
the parameter values described in section 4.1. 

Fig. 2a The proton charge form factor 
of the funnel model (solid line) compared with data \cite {Hoe}.
The dash-dot and dash lines show the
form factors of the funnel and oscillator models with
point-quark charge distributions.

Fig. 2b  The neutron charge form factor  of the funnel model
(solid line) in comparison with an empirical fit \cite{Hoe}
(dashed line).

Fig. 3 Proton and neutron  magnetic form factors of 
the funnel model compared to empirical fits (dashed lines).
 \cite{Hoe} 

\end {document}
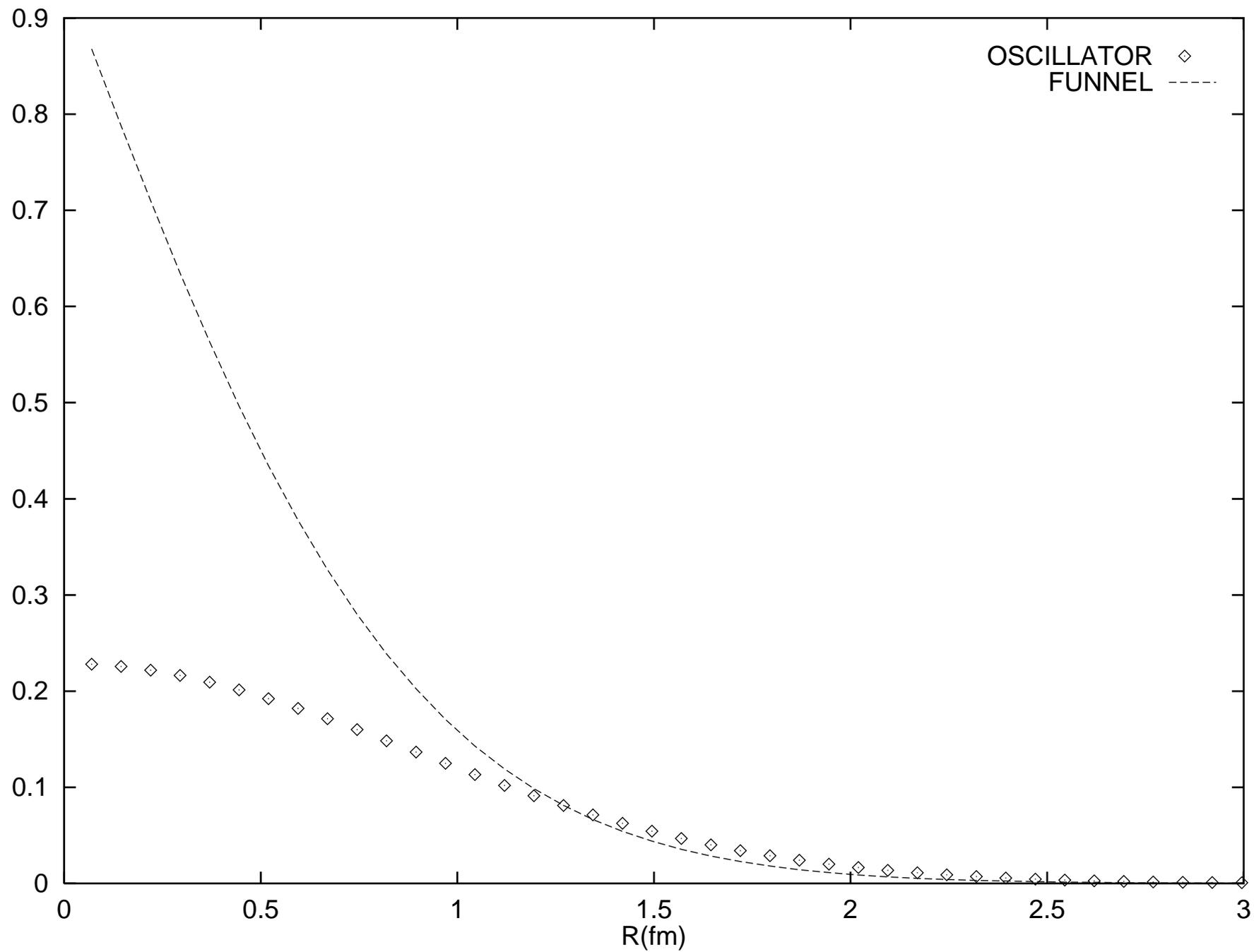
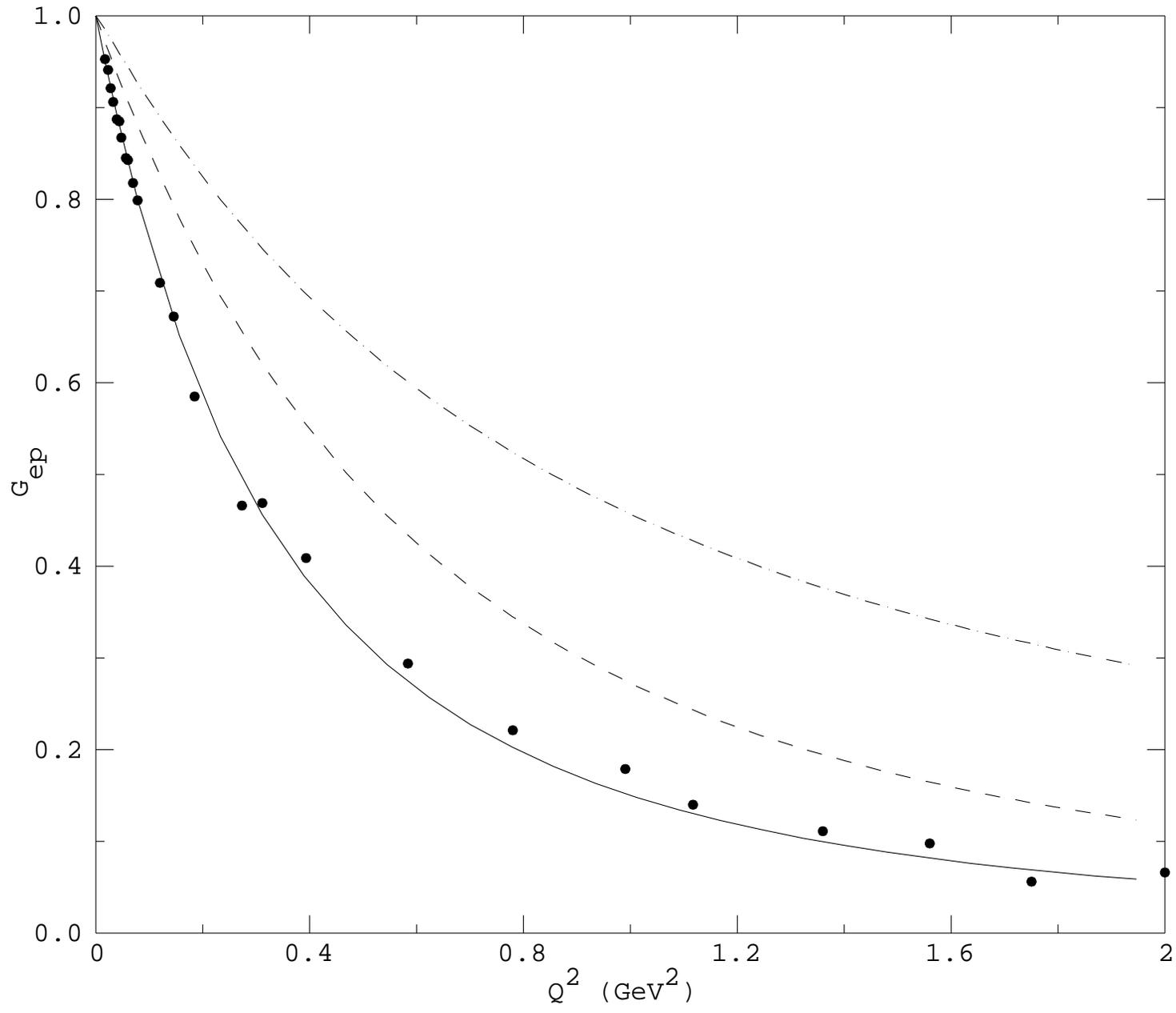
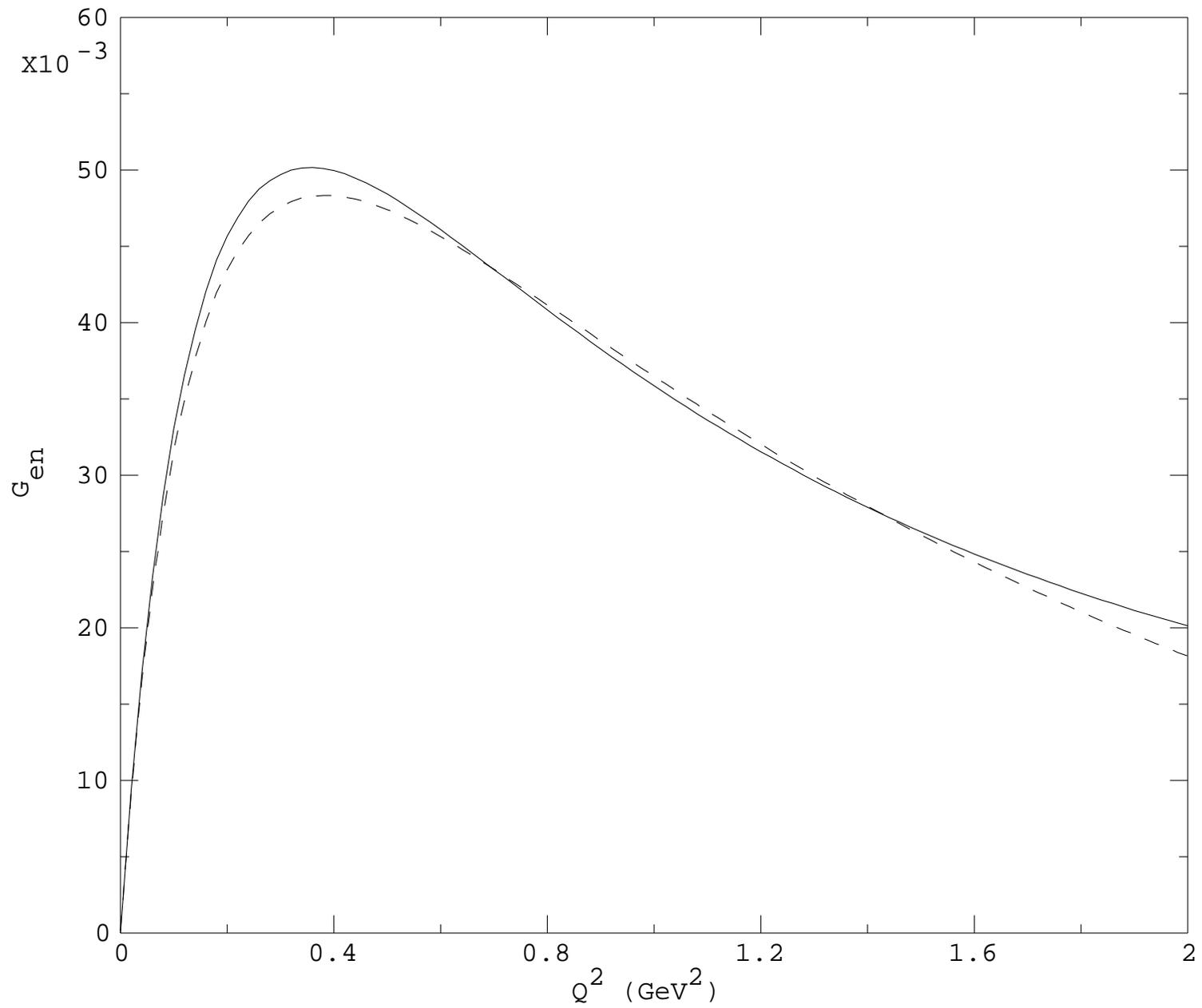
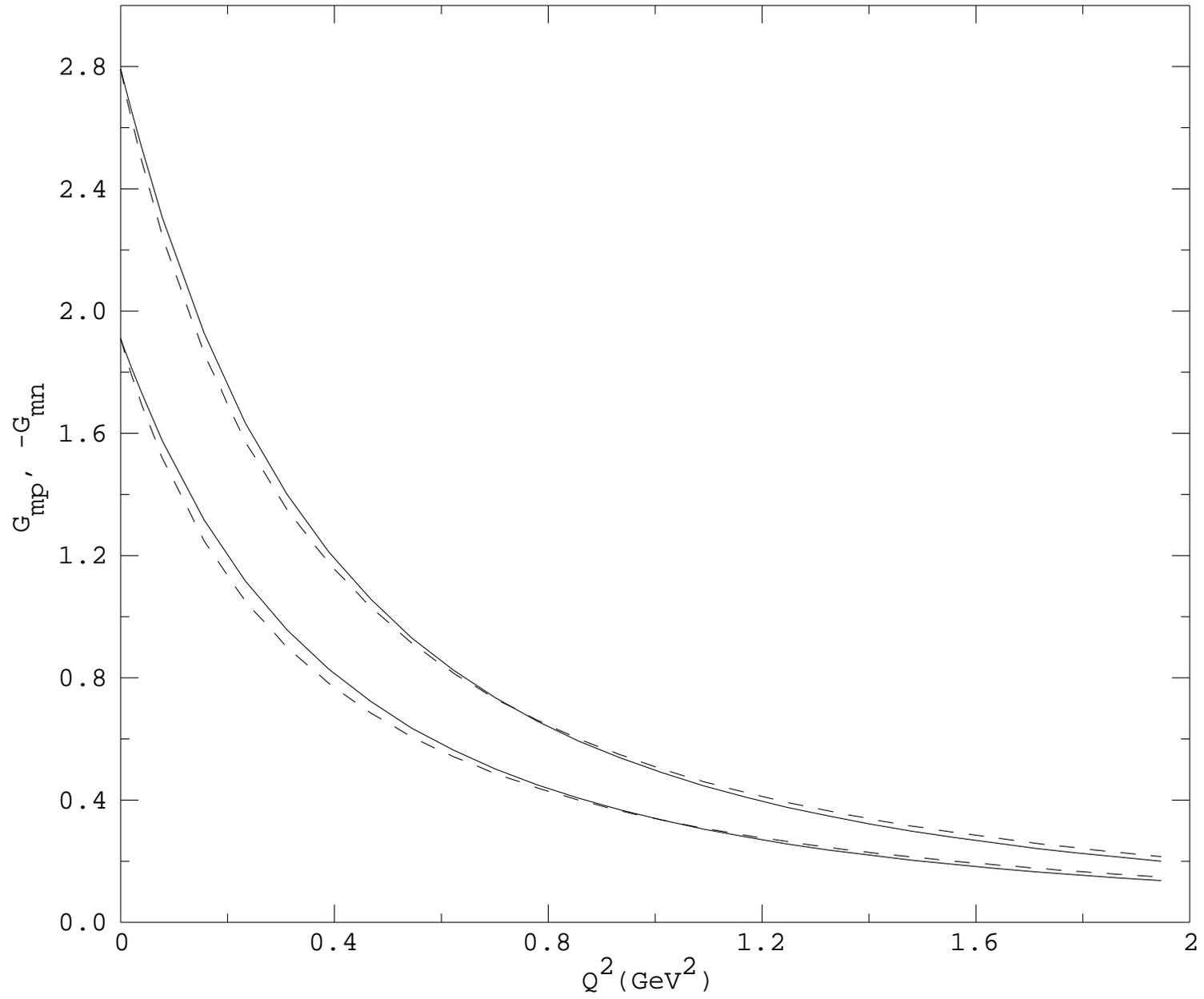
